\documentclass[fleqn,usenatbib]{mnras}

\usepackage{newtxtext,newtxmath}

\usepackage[T1]{fontenc}

\DeclareRobustCommand{\VAN}[3]{#2}
\let\VANthebibliography\thebibliography
\def\thebibliography{\DeclareRobustCommand{\VAN}[3]{##3}\VANthebibliography}

\usepackage{graphicx}	
\usepackage{amsmath}	
\usepackage{amssymb}	

\usepackage[usenames]{color} 
\usepackage{color}

\title[Triply eclipsing triple star TIC\,278825952]{TIC\,278825952: a triply eclipsing hierarchical triple system with the most intrinsically circular outer orbit}

\author[T. Mitnyan et al.]{
T. Mitnyan,$^{1}$\thanks{E-mail: mtibor@titan.physx.u-szeged.hu},
T. Borkovits$^{1,2,3}$,
S. A. Rappaport$^4$,
A. P\'al$^{2,5}$,
P.~F.~L.~Maxted$^{6}$
\\
$^{1}$Baja Astronomical Observatory of University of Szeged, H-6500 Baja, Szegedi \'ut, Kt. 766, Hungary\\
$^{2}$Konkoly Observatory, Research Centre for Astronomy and Earth Sciences,  H-1121 Budapest, Konkoly Thege Mikl\'os \'ut 15-17, Hungary\\
$^3$ ELTE Gothard Astrophysical Observatory, H-9700 Szombathely, Szent Imre h. u. 112, Hungary \\
$^4$ Department of Physics, Kavli Institute for Astrophysics and Space Research, M.I.T., Cambridge, MA 02139, USA\\
$^5$ ELTE E\"otv\"os Lor\'and University, Institute of Physics, H-1117 Budapest, P\'azm\'any P\'eter stny. 1A Hungary \\
$^6$ Astrophysics Group, Keele University, Staffordshire, ST5 5BG, UK\\
}

\date{Accepted XXX. Received YYY; in original form ZZZ}

\pubyear{2020}

\begin{document}
\label{firstpage}
\pagerange{\pageref{firstpage}--\pageref{lastpage}}
\maketitle

\begin{abstract}
We report the discovery of a compact triply eclipsing triple star system in the southern continuous viewing zone of the \textit{TESS} space telescope. TIC\,278825952 is a previously unstudied, circular eclipsing binary with a period of 4.781 days with a tertiary component in a wider, circular orbit of 235.55 days period that was found from three sets of third-body eclipses and from light travel-time effect dominated eclipse timing variations. We performed a joint photodynamical analysis of the eclipse timing variation curves, photometric data, and the spectral energy distribution, coupled with the use of \texttt{PARSEC} stellar isochrones.  We find that the inner binary consists of slightly evolved, near twin stars of masses of 1.12 and 1.09 $M_\odot$ and radii of 1.40 and 1.31 $R_\odot$. The third, less massive star has a mass of 0.75 $M_\odot$ and radius of 0.70 $R_\odot$. The low mutual inclination and eccentricities of the orbits show that the system is highly coplanar and surprisingly circular.

\end{abstract}




\section{Introduction}

 Triply eclipsing hierarchical triple systems are not easy to spot. The reason for this is that their discoveries need not only a lucky orbital configuration of the stars, but also long-term, continuous and precise photometric observations are essential to detect them. Thanks to the rapidly growing number of available light curves with such advantages produced by space-based photometric surveys (e.~g. \textit{CoRoT}, \textit{Kepler}, \textit{K2}, \textit{TESS}) their numbers are slowly, but steadily increasing.

These systems are very important in the field of stellar parameter determination, because they allow us to derive the orbital and physical parameters of all of the constituent stars (e.g., their individual masses) with a high precision. Moreover, with the appropriate data set, e.g., containing both light and radial velocity curves, this can be accomplished in a completely model-independent way. Almost all theoretical fields of stellar astronomy benefit from these precise parameters derived from observations. They can be used, e.g., to fine-tune star formation and stellar evolution theories or to analyze short- and long-term dynamical evolution of stellar systems \citep[see, e.~g.][]{2014ApJ...781L..13T,2014MNRAS.438.1909D,2018ApJ...854...44M}.

Until now, less than two dozen triply eclipsing hierarchical triple stellar systems are known in the literature (see an informative collection of them in \citealt{borkovits2020b}). While, in order to observe both inner and outer eclipses in a hierarchical triple system, it is necessary to view both the inner and outer orbits almost edge-on, there are no  restrictions (at least in theory) on the mutual inclination of the inner and outer orbital planes.\footnote{In this regard, note that KIC\,2835289, one of the 17 systems listed in Table~1 of \citet{borkovits2020b}, strictly speaking, is not a triply eclipsing triple, since in that case the inner binary  does not exhibit eclipses, only ellipsoidal variations} \citep[see][]{2014AJ....147...45C}. It is, of course, another question as to how the orbital plane precession, which necessarily occurs for non-coplanar systems, affects the visibility intervals of third-body eclipses, or even inner binary events, thereby providing a chance to detect such systems. This problem, in the context of transiting circumbinary planets was discussed in the papers of \citet{2015MNRAS.449..781M,2017MNRAS.465.3235M}. 

Regarding the above mentioned relative, or mutual, inclination of the inner and outer orbits, it is one of the most relevant parameters of a triple, or multiple stellar system from the point of view of its formation, and past and future evolutionary history \citep[see, e.~g.][for recent reviews]{2016ComAC...3....6T,2020A&A...640A..16T}. In this context, flat (coplanar) systems might have extraordinary importance. The reason is that for most of the hierarchical triple stars, the mutual inclination is subject to substantial variations (including even flip-flops from prograde to retrograde configurations and vice versa) due to various dynamical effects. A partial list of such effects includes the (eccentric) Lidov-Kozai effect \citep[see][for a review]{2016ARA&A..54..441N}, its interplay with tidal friction \citep[see, e.~g.][]{1998MNRAS.300..292K,2014ApJ...793..137N} (which may freeze the mutual inclination), and  resonant interactions with stellar spins \citep{2016CeMDA.126..189C}. In the case of a flat system, however, especially, as far as none of its components have undergone a Roche-lobe filling stage, one can expect that the alignment of the orbital planes is a certain relic of the formation process of the given triple star \citep[in this regard see, e.~g.][and references therein]{2020MNRAS.491.5158T}. 

The subset of compact, extremely flat systems with very accurately known mutual inclinations is very small, and listed in Table~5 of \citet{borkovits2020b}. More than half of these systems (four of seven) come from the triply eclipsing triples, emphasizing the extremely fortuitous role of such systems in determining accurate mutual inclinations. Among these objects, only one (HD\,181068, \citealt{2011Sci...332..216D}, \citealt{borko13}) has a highly circular outer orbit which is an unusual configuration for such  a system.

In this paper, we report the discovery and the first study of TIC\,278825952, a triply eclipsing compact, hierarchical triple stellar system which joins the company of HD\,181068 with a surprisingly circular outer orbit. The system is located in the southern continuous viewing zone (SCVZ) of the \textit{TESS} spacecraft and, therefore, was observed nearly continuously during Year 1 of the ongoing \textit{TESS} mission. This high precision and almost uninterrupted data set was essential in order to discover the triply eclipsing nature of the object and to determine its astrophysical and orbital parameters precisely. In Sect.~\ref{sec:obsdata} we describe all the available observational data and their preparation for the complex, joint photodynamical analysis which is discussed in Sect.~\ref{Sect:photodyn}. Then, the results are discussed and finally, summarized in Sects.~\ref{sec:discussion} and \ref{Sect:Summary}.

\section{Observational data}
\label{sec:obsdata}
\subsection{Catalog data}

In Table~\ref{tab:catalogs}, we collected photometric passband magnitudes of the system from different surveys, e.g. APASS \citep[AAVSO Photometric All Sky Survey;][]{munari2014}, 2MASS \citep[wo Micron All-Sky Survey;][]{cutri2003}, AllWISE \citep[Wide-field Infrared Survey Explorer: All-Sky Data Release;][]{cutri2014} and Gaia in order to construct the spectral energy distribution (`SED') of the system. The SED along with theoretical isochrones and the photodynamical model of the system provides an opportunity to determine the masses of the components in a model-dependent way (see Sect.~\ref{Sect:photodyn} for details). 

We found only two dedicated spectroscopic surveys containing data about TIC\,278825952 (RAVE DR5; Radial Velocity Experiment, \citealt{kunder2017}; TESS-HERMES DR1, \citealt{sharma2018}), despite the fact that this is a relatively bright system. These catalogs list spectroscopically determined effective temperature, $\log g$ and metallicity values assuming that the two stars in the binary dominate the light and are near twins. We can use these values to compare with our results from photometry, and thus we have included these quantities in Table~\ref{tab:catalogs} as well. We also note that the Gaia DR2 catalog lists a large RMS scatter for its radial velocity measurements (from 11 spectra), directly indicating the binary or, multiple, nature of the source.

\begin{table}
	\centering
	\caption{Main properties of TIC\,278825952 from different catalogs.}
	\label{tab:catalogs}
	\begin{tabular}{c c c}
		\hline
		Parameter & Value & References\\
		\hline
		RA [$\degr$]& 100.47064 & 1\\
		DEC [$\degr$] & -55.79494 & 1\\
		$\mu_{\mathrm{RA}}$ [mas\,yr$^{-1}$] & 1.16 $\pm$ 0.06 & 1\\
		$\mu_{\mathrm{DEC}}$ [mas\,yr$^{-1}$] & 13.52 $\pm$ 0.05 & 1\\
		$G$ [mag] & 11.8484 $\pm$ 0.0004 & 1\\
		$G_{\mathrm{BP}}$ [mag] & 12.147 $\pm$ 0.010 & 1\\
		$G_{\mathrm{R}P}$ [mag] & 11.401 $\pm$ 0.019 & 1\\
		$T$ [mag] & 11.457 $\pm$ 0.006 & 2\\
		B [mag] & 12.611 $\pm$ 0.172 & 3\\
		V [mag] & 12.062 $\pm$ 0.168 & 3\\
		g$'$ [mag] & 12.284 $\pm$ 0.175 & 3\\
		r$'$ [mag] & 11.911 $\pm$ 0.182 & 3\\
		i$'$ [mag] & 11.893 $\pm$ 0.269 & 3\\
		J [mag] & 10.897 $\pm$ 0.026 & 4\\
		H [mag] & 10.617 $\pm$ 0.024 & 4\\
		K [mag] & 10.526 $\pm$ 0.020 & 4\\
		W1 [mag] & 10.529 $\pm$ 0.023 & 5\\
		W2 [mag] & 10.547 $\pm$ 0.019 & 5\\
		Distance [pc] & 561 $\pm$ 8 & 6\\
		$T_{\mathrm{eff,\,RAVE}}$ [K] & 6175 $\pm$ 83 & 7\\
		$\log g_{\mathrm{\,RAVE}}$ [dex] & 4.19 $\pm$ 0.16 & 7\\
		$[M/H]_{\mathrm{\,RAVE}}$ [dex] & -0.32 $\pm$ 0.12 & 7\\
		$T_{\mathrm{eff,\,TESS-HERMES}}$ [K] & 6202 $\pm$ 120 & 8\\
		$\log g_{\mathrm{\,TESS-HERMES}}$ [dex] & 4.31 $\pm$ 0.20 & 8\\
		$[M/H]_{\mathrm{\,TESS-HERMES}}$ [dex] & -0.38 $\pm$ 0.10 & 8\\
		\hline
	\end{tabular}
	
\textbf{References. }(1) Gaia DR2 \citep{gaia2018}; (2) TIC-8 catalog \citep{stassun2018}; (3) APASS Landolt-Sloan BVgri Photometry of RAVE Stars. I. \citep{munari2014}; (4) 2MASS All-Sky Catalog of Point Sources \citep{cutri2003}; (5) AllWISE catalog \citep{cutri2014}; (6) \citet{bailer-jones2018}; (7) RAVE DR5 \citep{kunder2017}; (8) TESS-HERMES DR1 \citep{sharma2018}
\end{table}

\subsection{\textit{TESS} photometry}
\label{sec:TESSphot} 

The \textit{TESS} space telescope \citep{ricker2015} is monitoring a significant fraction of the sky, spending about a month on each $22\degr \times 96\degr$ sector of the sky. Nearly a whole hemisphere is thereby covered in a year. Regions close to the ecliptic poles are observed almost continuously, resulting in a year long data set for objects located in these regions. TIC\,278825952 has been observed during the first 13 sectors of \textit{TESS} with a 30-minute cadence on the Full-Frame Images (FFIs). However, data from Sector 7 is missing because our target fell between two CCDs due to its unfortunate positioning. Nevertheless, we still have a 1-year long continuous photometric data with a 1-month long gap in the middle which is more than suitable for further analysis of the system.

\begin{figure}
\begin{center}
\includegraphics[width=0.47 \textwidth]{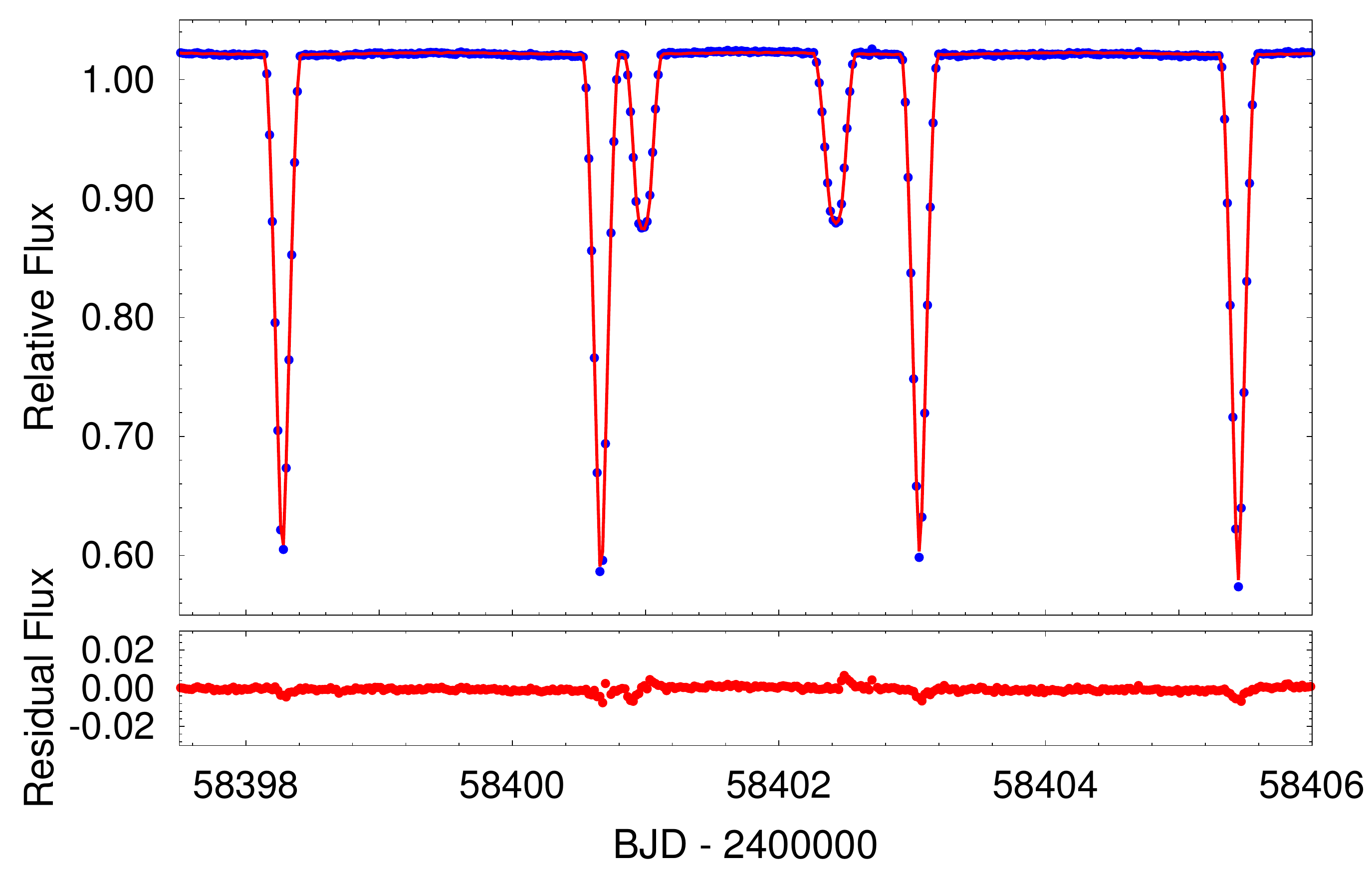}
\includegraphics[width=0.47 \textwidth]{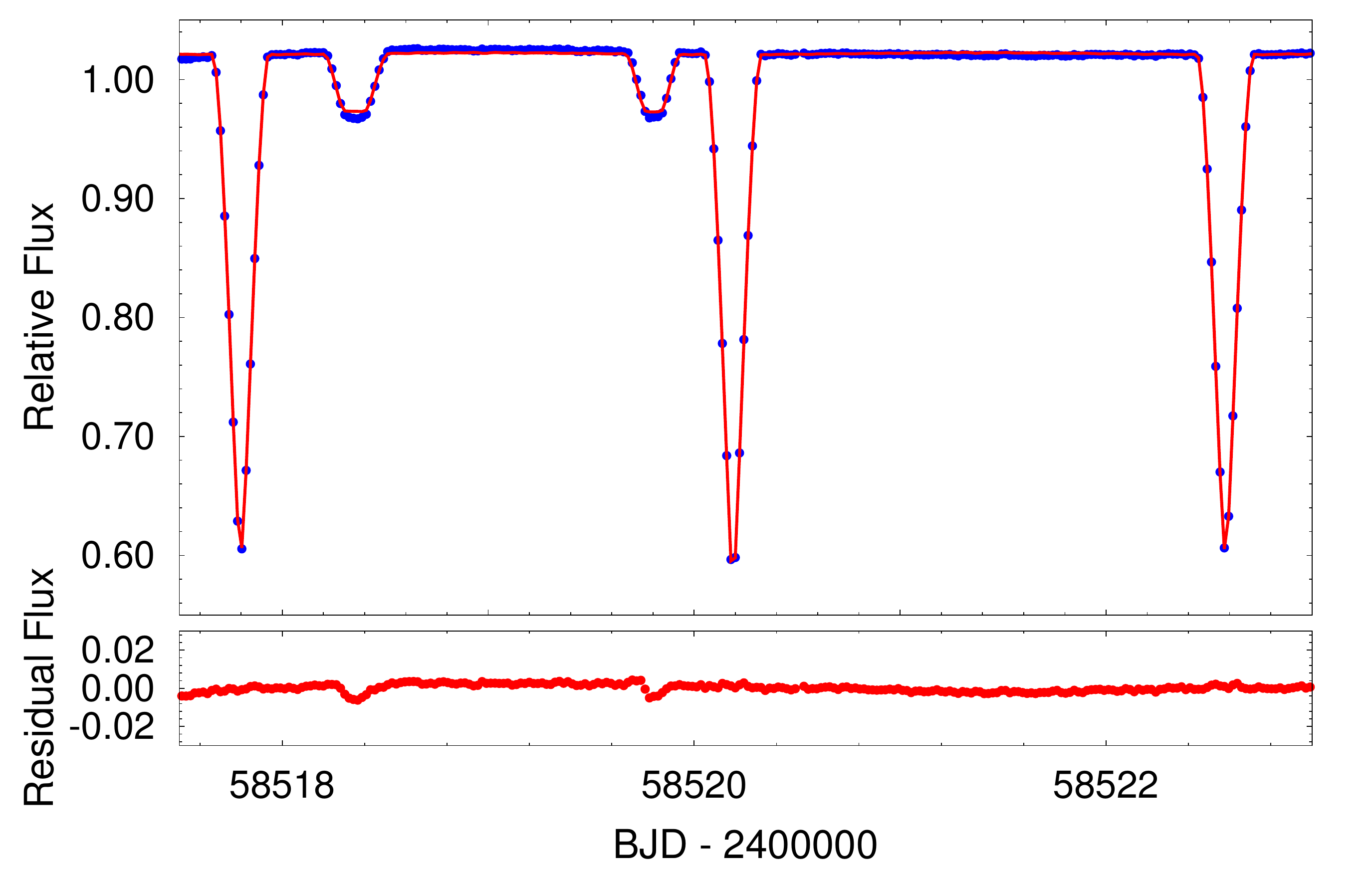}
\includegraphics[width=0.47 \textwidth]{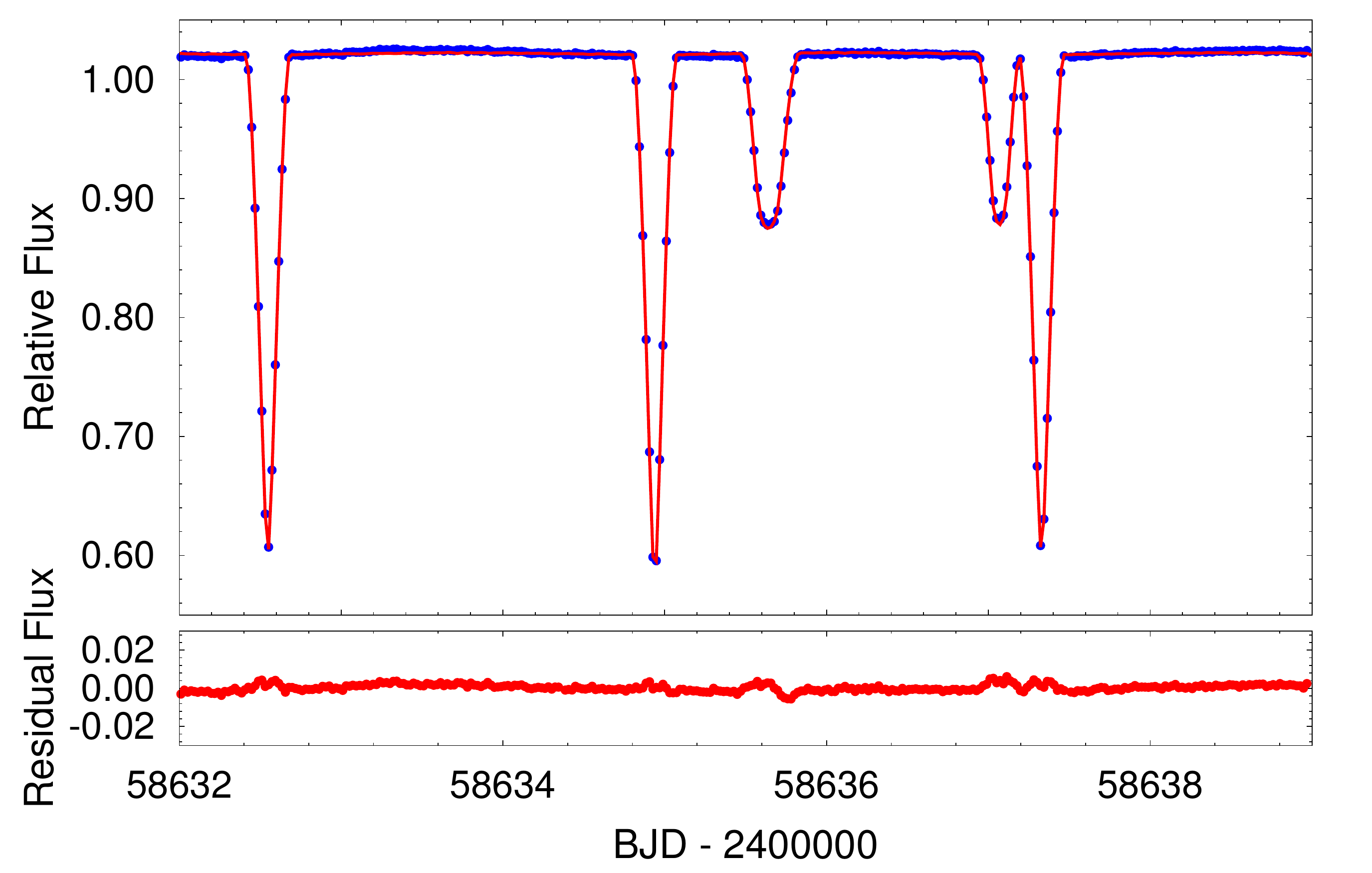}
\caption{Three sets of third-body eclipses. The blue circles represent \textit{TESS} observations. The red curve is the photodynamical model solution corrected for the 30-minute integration time (see later, in Sect.~\ref{Sect:photodyn}); the residuals to the model are also shown below the light curves. The top and bottom panels show events where the less massive distant companion passes separately in front of the two members of the inner binary. The middle panel displays the two anomalous eclipses when the two components of the inner pair passes in front of the third star.} 
\label{fig:outecl} 
\end{center}
\end{figure}

After downloading the FFI data from the MAST portal\footnote{\href{https://mast.stsci.edu}{https://mast.stsci.edu}}, the light curve from each sector was obtained by a convolution-based differential photometric pipeline based on the {\sc Fitsh} package \citep{pal2012} -- see also \citet{borkovits2020a} for more details about the practical implementation. For a final step, we converted the magnitudes to normalized fluxes and applied a careful detrending method using the {\sc W{\={o}}tan} package \citep{hippke2019} which filters out any instrumental trends and preserves all other features in the light curves. Sections of the detrended light curve of TIC\,278825952 based on all available \textit{TESS} observations can be seen in Fig.~\ref{fig:outecl}. It shows primary and secondary eclipses of the inner binary stars with an orbital period of $4.78$\,d. The light curve also reveals three pairs of extra eclipses (in Sectors  3, 8, and 12) caused by the third star on a wider (`outer') orbit passing in front of and behind the inner binary components. This immediately shows that the target is at least a hierarchical triple star with an outer orbital period of $235.5$\,d.

\subsection{WASP photometry}

TIC\,278825952 was also observed in the images of the WASP-South project \citep{2006PASP..118.1407P,2006MNRAS.373..799C} during four seasons between September 2008 and March 2012. The WASP instruments each consists of an array of 8 cameras with Canon 200-mm f/1.8 lenses and  2k$\times$2k $e$2$V$ CCD detectors providing images with a field-of-view  of $7.8\degr\times 7.8\degr$ at an image scale of 13.7 arcsec/pixel. Images are obtained through a broad-band filter covering 400-700\,nm. Fluxes are measured in an aperture with a radius of 48 arcsec for the WASP data. The data are processed with the SYSRem algorithm \citep{2005MNRAS.356.1466T} to remove instrumental effects. 

\begin{figure}
\begin{center}
\includegraphics[width=0.47\textwidth]{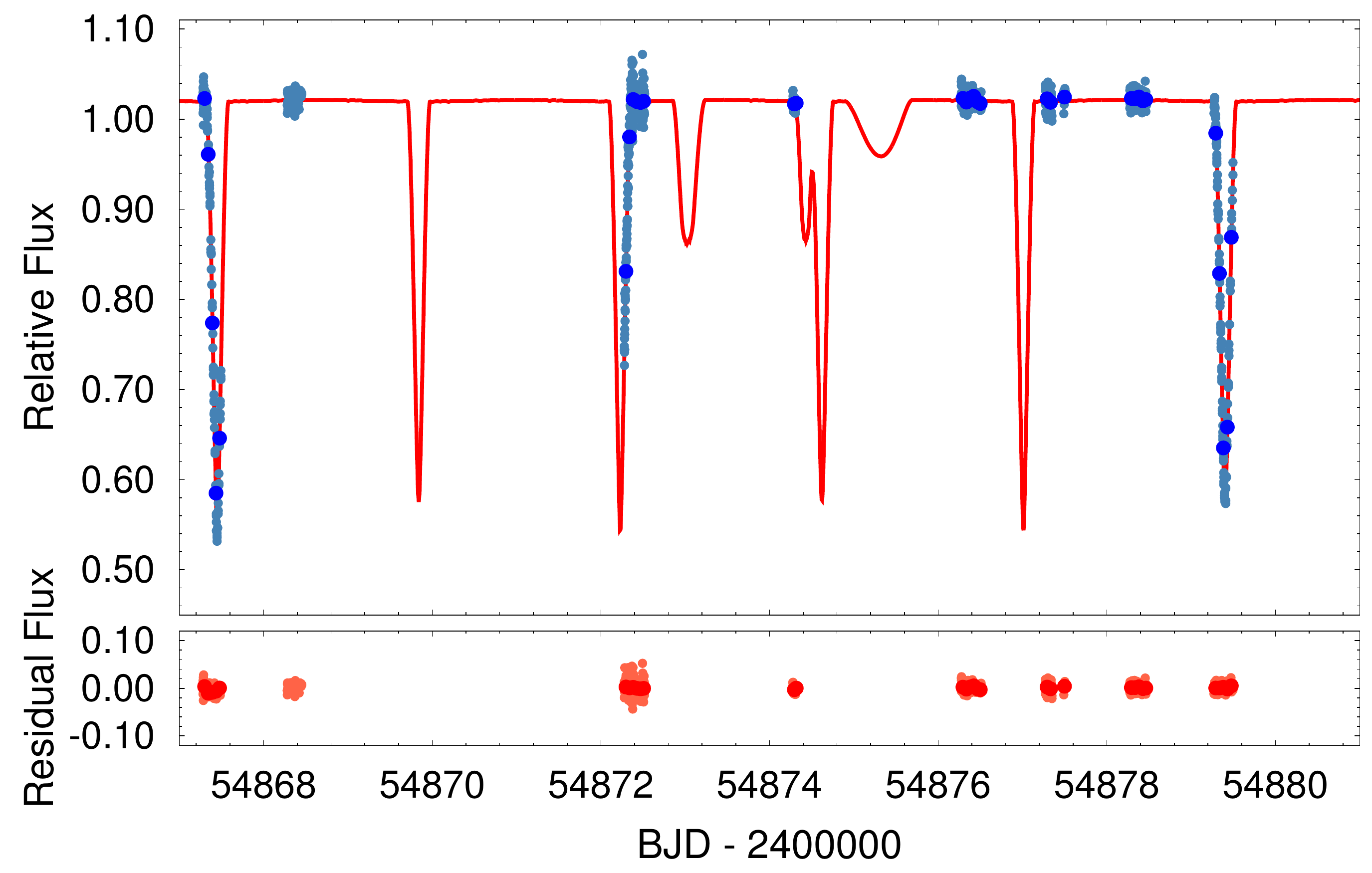}
 \includegraphics[width=0.47\textwidth]{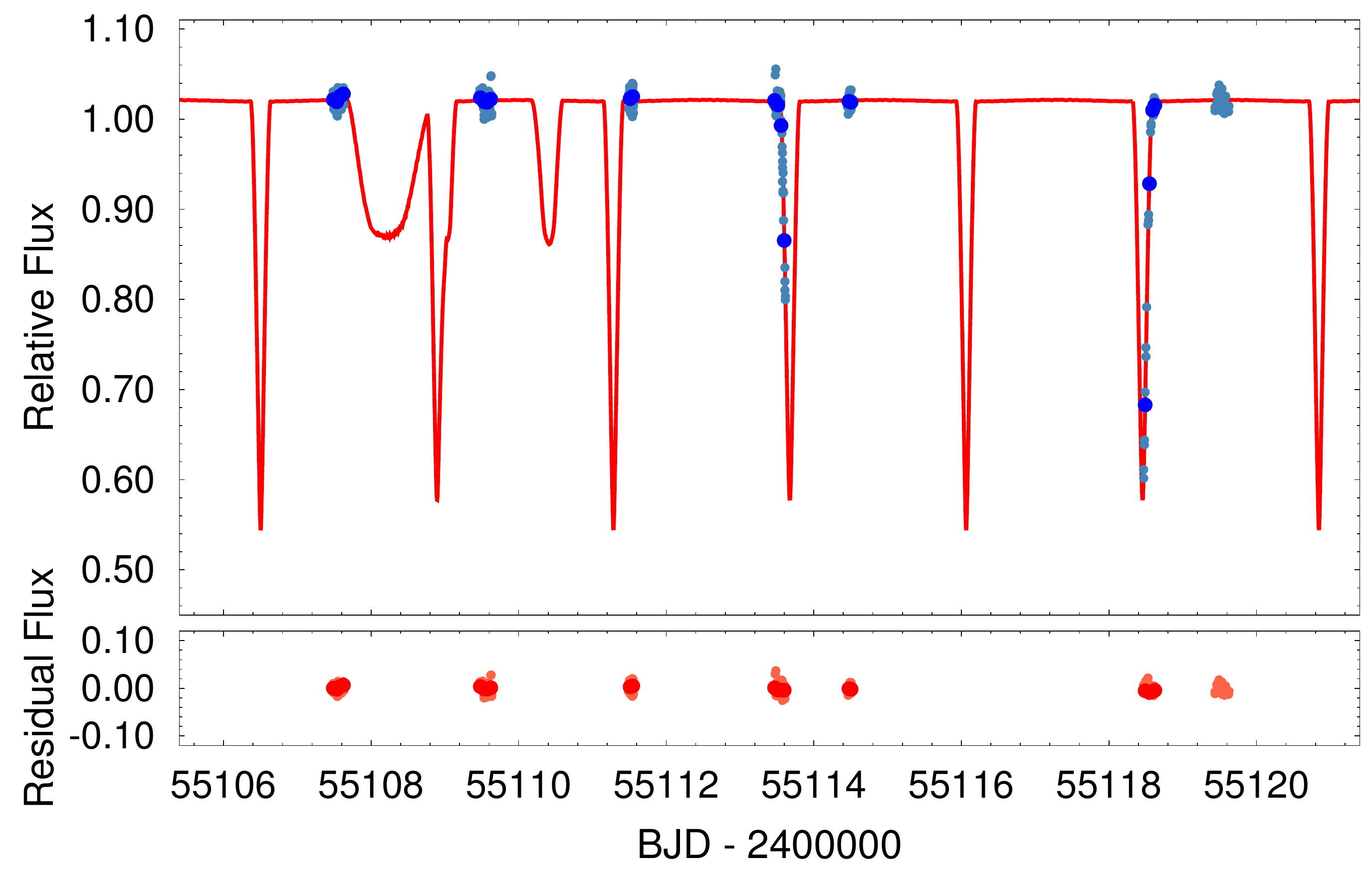}
\caption{Two sections of WASP observations for TIC\,278825952. The pale and dark blue points are the original and the 1-hr averaged WASP photometric data for this target, respectively.  Only the dark blue points were used in the fit. The solid red curve is the photodynamical model.}
\label{fig:wasp_lc} 
\end{center}
\end{figure}

While archival WASP data have lower quality than the high-accuracy \textit{TESS} observations, and are also subjected to diurnal and seasonal data gaps, their use was essential for, and therefore they were included in, our analyses (see below in Sect.~\ref{Sect:photodyn}). Sections of WASP data are plotted in Fig.~\ref{fig:wasp_lc}.

\subsection{ASAS-SN photometry}

\begin{figure}
\begin{center}
\includegraphics[width=0.47\textwidth]{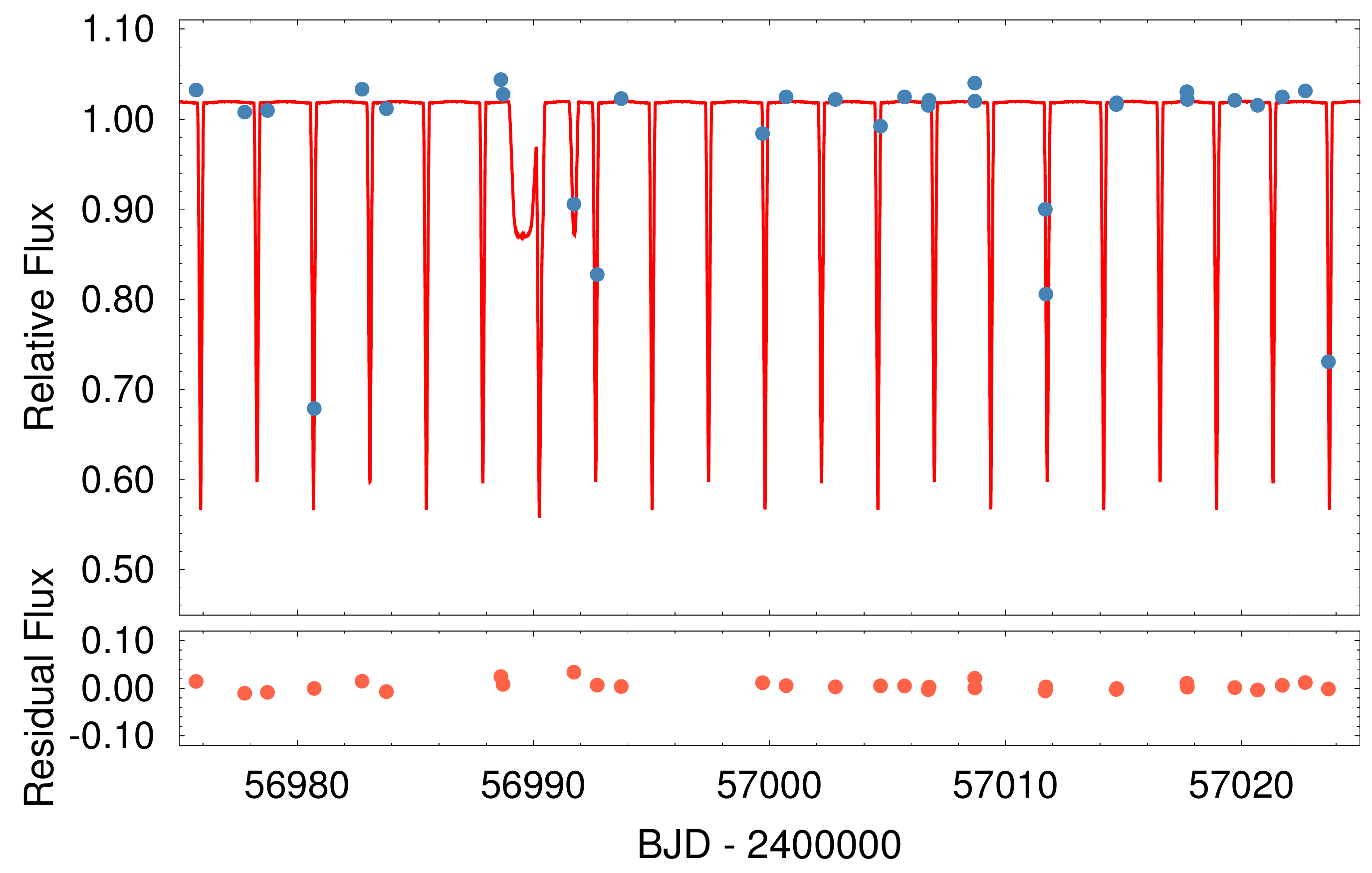}
\caption{A 50-day-long section of ASAS-SN observations of TIC\,278825952. The pale blue points are the original $V$-band observations converted to normalized flux values (and also times to BJD). The solid red curve is the photodynamical model calculated for the appropriate dates. (The ASAS-SN data themselves were not included into the complex photodynamical modelling.) The bottom panel shows the residuals to the photodynamical model.}
\label{fig:asassn_lc} 
\end{center}
\end{figure}

ASAS-SN\footnote{\href{http://www.astronomy.ohio-state.edu/asassn/index.shtml}{http://www.astronomy.ohio-state.edu/asassn/index.shtml}} (All-Sky Automated Survey for Supernovae) currently has 24 telescopes around the Earth covering the entire sky. TIC\,278825952 was included in a targeted, supplementary survey towards the SCVZ of \textit{TESS} conducted by the ASAS-SN network \citep{2019MNRAS.485..961J}. For these observations the units named ``Brutus” (Haleakala, Hawaii) and ``Cassius” (CTIO, Chile) were used which are comprised of four 14-cm telescopes each with a field of view of 4.5\,deg$^2$ and a pixel size of 8$''$. The images were processed using image subtraction combined with aperture photometry, and the resultant light curves are publicly available in the ASAS-SN light curve server\footnote{\href{https://asas-sn.osu.edu/}{https://asas-sn.osu.edu/}} \citep{2014ApJ...788...48S,2017PASP..129j4502K}. According to the back-projected photodynamical model (see Sect.~\ref{Sect:photodyn}, below) two ASAS-SN photometry points belong to `extra' third-body eclipsing events. One of them is shown in that 50-day-long section of the ASAS-SN observations which is plotted in Fig.~\ref{fig:asassn_lc}.

\subsection{ETV data}

We determined mid-eclipse times of the regular eclipses of the inner binary pair from both the \textit{TESS} and WASP data in the manner described by \citet{borko16}. For the inner binary we found the following linear ephemeris:
\begin{equation}
    \mathrm{MIN}_I \mathrm{[BJD]}=2458328.9473+4\fd7810765\times E,
    \label{eq:ETV_ephem}
\end{equation}
using the resultant eclipse timing variation (ETV) curve where $E$ denotes the cycle number (integer and half-integer for primary and secondary eclipses, respectively). The eclipse times are tabulated in Table~\ref{etvtab}. 
 
\begin{table*}
\caption{Times of minima of TIC\,278825952.}
 \label{etvtab}
\begin{tabular}{@{}lrllrllrl}
\hline
BJD & Cycle  & std. dev. & BJD & Cycle  & std. dev. & BJD & Cycle  & std. dev. \\ 
$-2\,400\,000$ & no. &   \multicolumn{1}{c}{$(d)$} & $-2\,400\,000$ & no. &   \multicolumn{1}{c}{$(d)$} & $-2\,400\,000$ & no. &   \multicolumn{1}{c}{$(d)$} \\ 
\hline
54757.482270 & -747.0 & 0.000387 & 58388.712235 &   12.5 & 0.000150 & 58558.438283 &   48.0 & 0.000130 \\ 
54793.337351$^*$& -739.5 & 0.020109 & 58391.102794 &   13.0 & 0.000149 & 58560.829015 &   48.5 & 0.000176 \\ 
54824.418762 & -733.0 & 0.000588 & 58393.493445 &   13.5 & 0.000167 & 58563.219460 &   49.0 & 0.000172 \\ 
54836.371648 & -730.5 & 0.000142 & 58395.883921 &   14.0 & 0.000168 & 58565.610072 &   49.5 & 0.000136 \\ 
54855.496260 & -726.5 & 0.000193 & 58398.274543 &   14.5 & 0.000141 & 58568.000771 &   50.0 & 0.000165 \\ 
54867.449100 & -724.0 & 0.000195 & 58403.055679 &   15.5 & 0.000117 & 58570.391562 &   50.5 & 0.000159 \\ 
54879.402087 & -721.5 & 0.000161 & 58405.446033 &   16.0 & 0.000178 & 58572.782019 &   51.0 & 0.000146 \\ 
54891.353369$^*$ & -719.0 & 0.000744 & 58407.836754 &   16.5 & 0.000158 & 58575.172544 &   51.5 & 0.000151 \\ 
54915.259506 & -714.0 & 0.000346 & 58412.617970 &   17.5 & 0.000126 & 58577.563250 &   52.0 & 0.000173 \\ 
55216.464593 & -651.0 & 0.000248 & 58415.008195 &   18.0 & 0.000184 & 58579.953796 &   52.5 & 0.000133 \\ 
55228.419356 & -648.5 & 0.000286 & 58417.398842 &   18.5 & 0.000148 & 58584.734955 &   53.5 & 0.000134 \\ 
55240.370789 & -646.0 & 0.000149 & 58422.179783 &   19.5 & 0.000152 & 58587.125764 &   54.0 & 0.000144 \\ 
55486.597328 & -594.5 & 0.001164 & 58426.960811 &   20.5 & 0.000135 & 58589.516421 &   54.5 & 0.000157 \\ 
55632.421040$^*$ & -564.0 & 0.005078 & 58429.351259 &   21.0 & 0.000191 & 58591.906868 &   55.0 & 0.000177 \\ 
55859.521341 & -516.5 & 0.000742 & 58431.741750 &   21.5 & 0.000158 & 58594.297521 &   55.5 & 0.000148 \\ 
55883.424951 & -511.5 & 0.000663 & 58434.131950 &   22.0 & 0.000164 & 58599.078853 &   56.5 & 0.000145 \\ 
55902.550569 & -507.5 & 0.001333 & 58436.522479 &   22.5 & 0.000159 & 58601.469374 &   57.0 & 0.000197 \\ 
55907.332211 & -506.5 & 0.002108 & 58438.912970 &   23.0 & 0.000175 & 58603.859911 &   57.5 & 0.000156 \\ 
55926.455406 & -502.5 & 0.000547 & 58441.303409 &   23.5 & 0.000121 & 58606.250831 &   58.0 & 0.000210 \\ 
55938.406971 & -500.0 & 0.000539 & 58443.693772 &   24.0 & 0.000192 & 58608.641388 &   58.5 & 0.000139 \\ 
55950.361557 & -497.5 & 0.001466 & 58446.084414 &   24.5 & 0.000138 & 58611.031836 &   59.0 & 0.000181 \\ 
55962.313877 & -495.0 & 0.000700 & 58448.474692 &   25.0 & 0.000183 & 58613.422325 &   59.5 & 0.000136 \\ 
55981.437680 & -491.0 & 0.000400 & 58453.255522 &   26.0 & 0.000222 & 58615.813062 &   60.0 & 0.000164 \\ 
55993.392108 & -488.5 & 0.000915 & 58455.646086 &   26.5 & 0.000129 & 58618.203678 &   60.5 & 0.000123 \\ 
56005.345204 & -486.0 & 0.000504 & 58458.036424 &   27.0 & 0.000131 & 58620.594258 &   61.0 & 0.000188 \\ 
58326.556287 &   -0.5 & 0.000132 & 58460.426973 &   27.5 & 0.000134 & 58622.984802 &   61.5 & 0.000144 \\ 
58328.946803 &    0.0 & 0.000153 & 58462.817170 &   28.0 & 0.000136 & 58625.375442 &   62.0 & 0.000202 \\ 
58331.337452 &    0.5 & 0.000141 & 58469.988543 &   29.5 & 0.000146 & 58627.766024 &   62.5 & 0.000117 \\ 
58333.728048 &    1.0 & 0.000161 & 58472.378923 &   30.0 & 0.000160 & 58630.156508 &   63.0 & 0.000173 \\ 
58336.118809 &    1.5 & 0.000148 & 58474.769537 &   30.5 & 0.000167 & 58632.547150 &   63.5 & 0.000127 \\ 
58340.899927 &    2.5 & 0.000135 & 58479.550257 &   31.5 & 0.000155 & 58634.937808 &   64.0 & 0.000175 \\ 
58343.290533 &    3.0 & 0.000179 & 58481.940766 &   32.0 & 0.000177 & 58642.109352 &   65.5 & 0.000143 \\ 
58345.681238 &    3.5 & 0.000131 & 58484.331260 &   32.5 & 0.000182 & 58644.499886 &   66.0 & 0.000184 \\ 
58348.071763 &    4.0 & 0.000235 & 58486.721705 &   33.0 & 0.000182 & 58646.890471 &   66.5 & 0.000155 \\ 
58350.462410 &    4.5 & 0.000161 & 58489.112072 &   33.5 & 0.000138 & 58649.280977 &   67.0 & 0.000145 \\ 
58352.853152 &    5.0 & 0.000163 & 58517.798008 &   39.5 & 0.000126 & 58651.671307 &   67.5 & 0.000134 \\ 
58355.243884 &    5.5 & 0.000187 & 58520.188724 &   40.0 & 0.000195 & 58654.062113 &   68.0 & 0.000160 \\ 
58357.634292 &    6.0 & 0.000152 & 58522.579331 &   40.5 & 0.000129 & 58656.452612 &   68.5 & 0.000125 \\ 
58360.024965 &    6.5 & 0.000158 & 58524.969818 &   41.0 & 0.000165 & 58658.842895 &   69.0 & 0.000159 \\ 
58362.415560 &    7.0 & 0.000200 & 58527.360298 &   41.5 & 0.000121 & 58661.233378 &   69.5 & 0.000131 \\ 
58364.806310 &    7.5 & 0.000120 & 58536.922685 &   43.5 & 0.000150 & 58663.623904 &   70.0 & 0.000150 \\ 
58369.587388 &    8.5 & 0.000157 & 58539.313401 &   44.0 & 0.000179 & 58666.014466 &   70.5 & 0.000125 \\ 
58371.978000 &    9.0 & 0.000183 & 58541.703979 &   44.5 & 0.000123 & 58670.795344 &   71.5 & 0.000141 \\ 
58374.368682 &    9.5 & 0.000158 & 58544.094661 &   45.0 & 0.000182 & 58673.185772 &   72.0 & 0.000177 \\ 
58376.759237 &   10.0 & 0.000177 & 58546.485117 &   45.5 & 0.000158 & 58675.576364 &   72.5 & 0.000152 \\ 
58379.149732 &   10.5 & 0.000143 & 58548.875749 &   46.0 & 0.000199 & 58677.966599 &   73.0 & 0.000158 \\ 
58383.931120 &   11.5 & 0.000120 & 58551.266533 &   46.5 & 0.000125 & 58680.357076 &   73.5 & 0.000152 \\ 
58386.321564 &   12.0 & 0.000158 & 58553.656951 &   47.0 & 0.000226 &&& \\ 
\hline
\end{tabular}

\textit{Notes.} Integer and half-integer cycle numbers refer to primary and secondary eclipses, respectively. Eclipses between cycle nos. $-747.0$ and $-486.0$ were observed in the WASP project. Other eclipse times were determined from the \textit{TESS} measurements. The eclipse times denoted by asterisks are considered to be outliers and were omitted from the analysis.   
\end{table*}

\begin{figure*}
\begin{center}
\includegraphics[width=0.99 \textwidth]{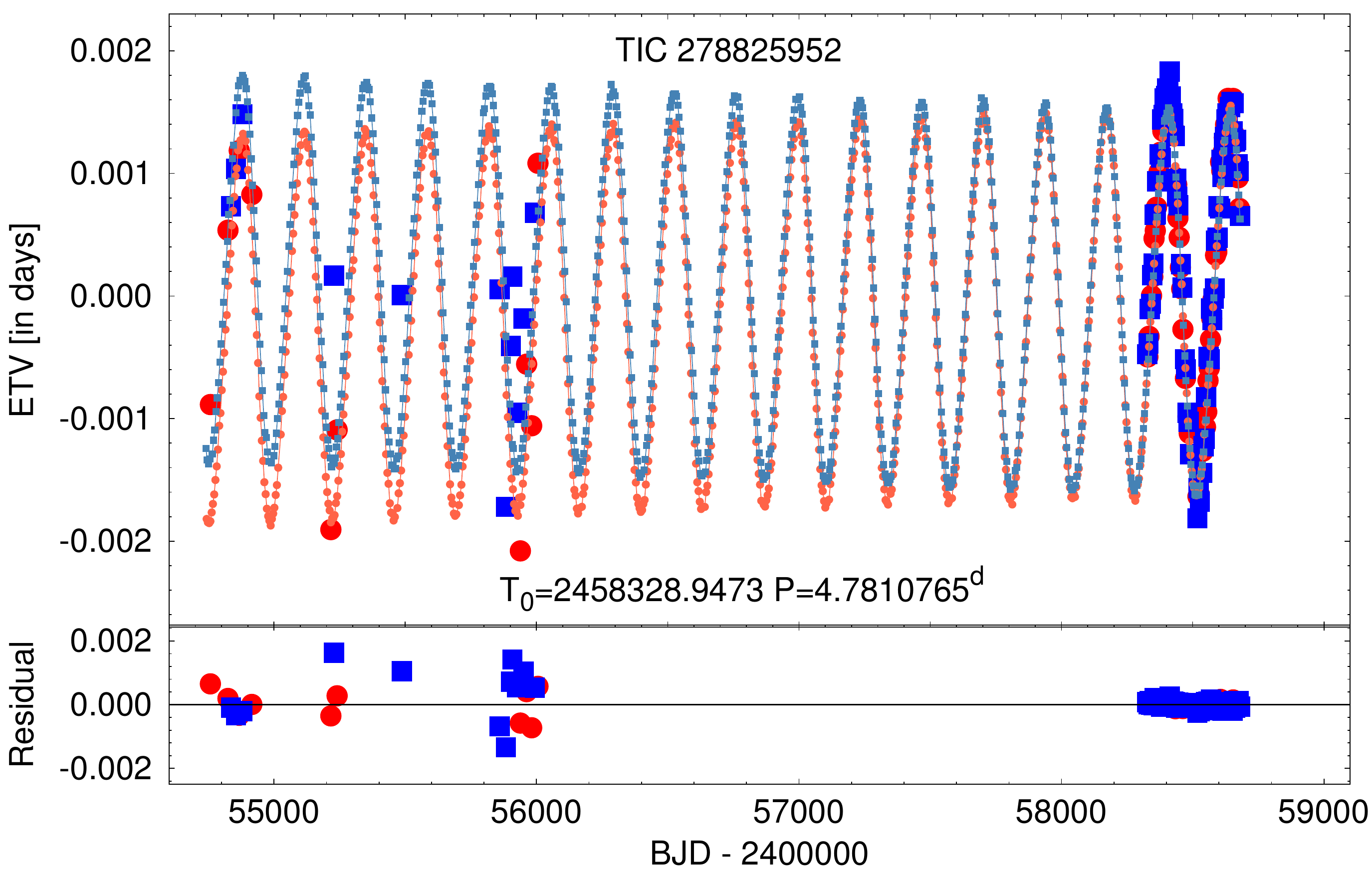}
\includegraphics[width=0.47 \textwidth]{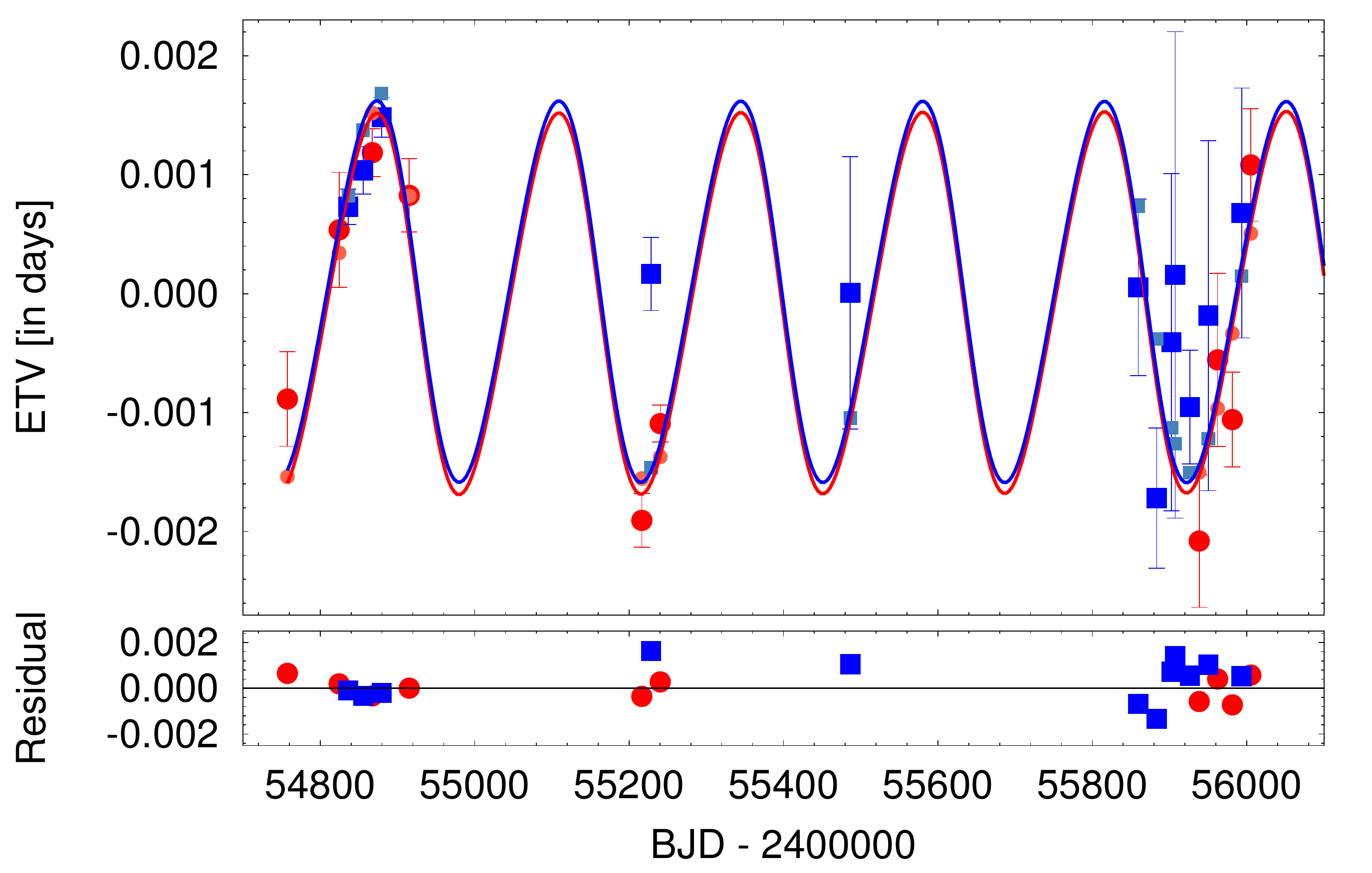}\includegraphics[width=0.47 \textwidth]{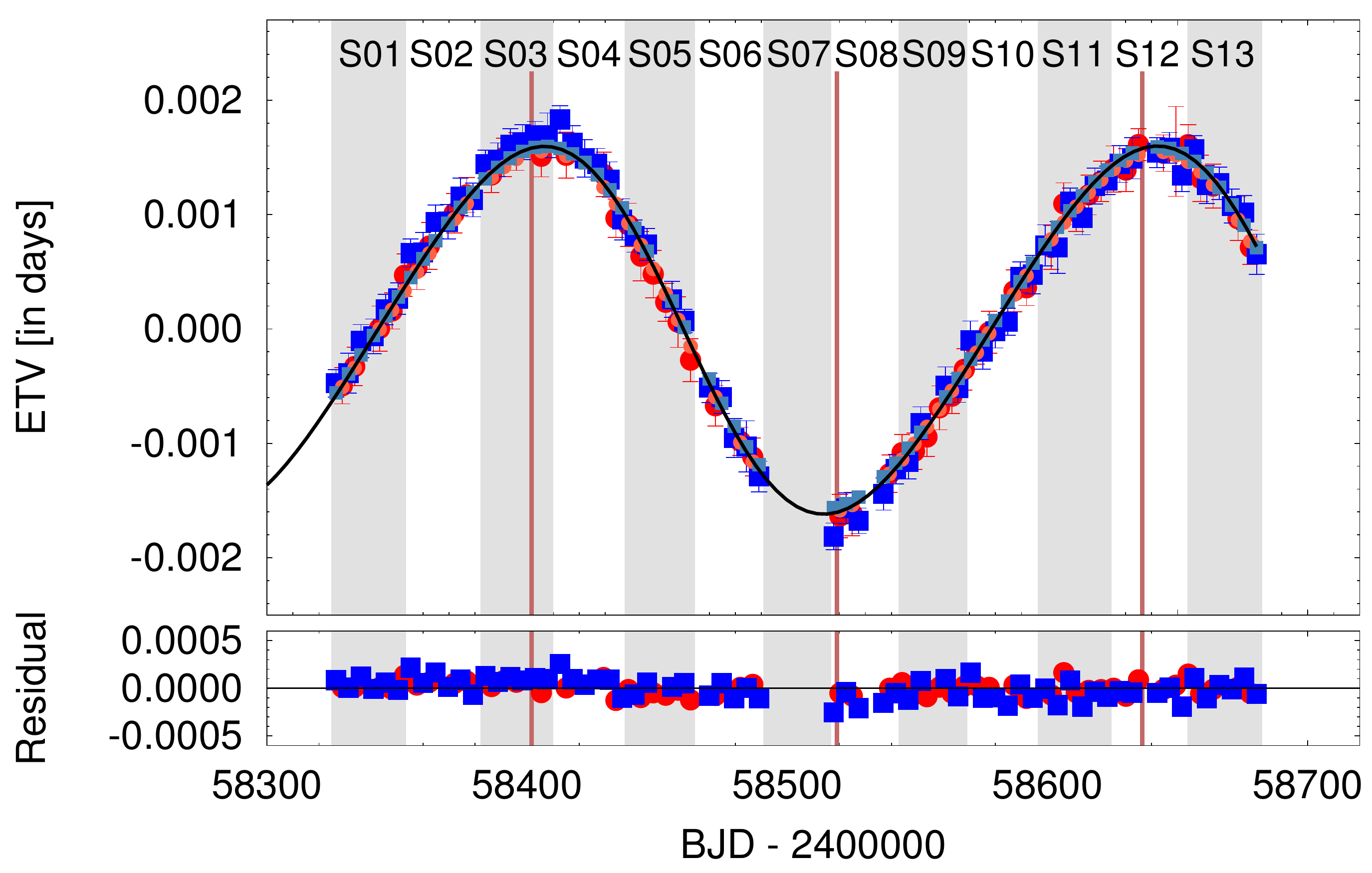}
\caption{Eclipse timing variations of TIC\,278825952. The large red filled circles and blue squares are calculated from the observed eclipse events, while the corresponding smaller symbols with lighter colors are determined from the photodynamical model solution. These model ETV points are connected to each other simply to guide the reader's eye.  For better visibility, bottom panels show zoom-ins of the ETV curves for the epochs of the WASP (lower left panel) and the \textit{TESS} observations (lower right panel). In these lower panels the continuous curves represent approximate analytic solutions obtained with the formulae of \citet{borko15} and, furthermore, in the lower right panel alternating grey and white stripes show the nominal time intervals of each sector during Year 1 of the \textit{TESS} mission, while the brown vertical lines mark the times of the third-body eclipses. The residuals of the observed vs. photodynamically modelled ETVs are plotted in the bottommost panels.} 
\label{etvs} 
\end{center}
\end{figure*} 

The overall ETV curve for TIC 278825952 is plotted in Fig.~\ref{etvs} along with the best-fit model that is described in the next section. The points determined both from the regular primary and secondary eclipses of the \textit{TESS} data (lower right panel of Fig.~\ref{etvs}) both exhibit the same closely sinusoidal shape curve, with $P_\mathrm{2}=235.5$\,days, revealing that both the inner and outer orbits are nearly circular. As one can see in the lower right panel of Fig.~\ref{etvs}, the outer eclipsing events closely correspond to the extrema of the ETV curve. Since in the present case the ETV curve is clearly light-travel-time-effect (LTTE) dominated (see Sect.~\ref{sec:discussion} for a discussion), this fact evidently shows that in the case of the deeper extra eclipses (in Sectors 3 and 12) the third star passes in front of the binary members, while in the case of the more shallow third-body events (in Sector 8) the situation is the opposite.

Turning to the older ETV points determined from the archival WASP data (lower left panel of Fig.~\ref{etvs}), they have significantly larger uncertainties. However, they show that no other periodic variations can be detected in the system, at least on the time scale of a decade.

\section{Joint analysis of the available data}
\label{Sect:photodyn}

We used the software package {\sc Lightcurvefactory} \citep[see][and further references therein]{borkovits2019,borkovits2020a,borkovits2020b} to carry out a complex photodynamical modeling of the system based on the data collected in Sect.~\ref{sec:obsdata}. The analysis has followed exactly the same steps which were discussed in \citet{,borkovits2020a,borkovits2020b} in detail, therefore there is no need to repeat it here. Instead, we note only some specific points.

For the photodynamical analysis, we fitted simultaneously (i) the 30-min cadence Year 1 \textit{TESS} light curve, (ii) the WASP light curve, (iii) the ETV curves of the primary and secondary eclipses and finally, (iv) the observed stellar SED in the form of catalogued passband magnitudes. As in our previous works, the SEDs were fitted to theoretical passband magnitudes\footnote{For the analysis we set the uncertainties of each measured passband magnitude to $\sigma=\mathrm{max}(0.03,\sigma_\mathrm{catalog})$ in order to take into account the intrinsic, systematic effects coming from the interpolation of the \texttt{PARSEC} grids and, furthermore to avoid the over-representation of the extremely accurate Gaia $G$ magnitude over the other data.} calculated with interpolation from the grids of tabulated theoretical \texttt{PARSEC} isochrones \citep{bressanetal12}.

In order to reduce computational costs we dropped out from the analysis the out-of-eclipse sections of both the \textit{TESS} and WASP light curves, i.e., in other words, only the $\phi=\pm0\fp05$ phase sections of the regular eclipses were kept, with the exception of the outer eclipses, where longer, 6-8-day-long sections of the light curves were also retained. Note that, in case of the WASP light curve, our treatment slightly departed from that which was followed in \citet{borkovits2020b}, where the full WASP light curve was considered in the photodynamical analysis. The reason is that, in the present situation, the \textit{TESS} light curve by itself made it possible to determine accurately the outer period and, therefore, we were able to pre-compute the expected locations of the outer eclipses in the intervals of the WASP observations. For further reduction in computational costs we formed 1-hr average points from the WASP observations, and these data were used for the analysis (naturally, with the appropriate cadence corrections).

Without radial velocity measurements the masses of the components cannot be determined directly; nevertheless, using theoretical \texttt{PARSEC} isochrones can help us to construct model-dependent masses for the components \citep[see][]{borkovits2020a}. This allowed us to constrain a physically and dynamically consistent model of the system and determine its orbital and physical properties. 

In the majority of the MCMC (Markov Chain Monte Carlo) runs the following parameters were adjusted:
\begin{itemize}
    \item [(i)] Nine of the twelve orbital element related parameters describing the two perturbed, osculating Keplerian orbits at epoch $t_0=2458320.0$, as follows: $e_1\cos\omega_1$, $e_1\sin\omega_1$, and $i_1$ giving the eccentricity, argument of periastron and the inclination of the inner orbit; furthermore, the parameters of the wide, outer orbit: $P_2$, $e_2\cos\omega_2$, $e_2\sin\omega_2$, $i_2$, the time of the inferior conjunction of the third component, $\mathcal{T}_2^{\mathrm{inf}}$, and the ascending node of the outer orbit, $\Omega_2$.\footnote{The angular orbital elements are defined in an observer related frame of which the base plane is the tangential plane of the sky. Furthermore, as $\Omega_1=0\degr$ was assumed at epoch $t_0$ for all runs, $\Omega_2$ set the initial trial value of the differences of the nodes ($\Delta\Omega$) which is the really relevant parameter for the modelling.}
    \item[(ii)] Three parameters connected to the stellar masses: primary star's mass, $m_\mathrm{A}$, and the mass ratios of the inner and outer subsystems $q_\mathrm{1,2}$.
    \item[(iii)] The passband-dependent extra lights $\ell_{\mathrm{TESS}}$, $\ell_{\mathrm{WASP}}$ account for two additional parameters.
    \item[(iv)] Finally, three parameters for the \texttt{PARSEC} isochrone and SED fitting: the logarithm of the age of the three stars, $\log\tau$, the metallicity [$M/H$] and the extinction coefficient $E(B-V)$.
\end{itemize}

Furthermore, twenty one additional parameters were internally constrained, as follows:
\begin{itemize}
    \item [(i)] The inner binary's orbital period, $P_1$, and the time of an inferior conjunction $\mathcal{T}_1^{\mathrm{inf}}$ of the secondary star of the inner pair at epoch $t_0$ were constrained via the ETV curves \citep[see appendix A of][]{borkovits2019}.
    \item[(ii)] The effective temperatures, $T_\mathrm{A,B,C}$, and radii, $R_\mathrm{A,B,C}$, of the three stars were calculated from interpolation at each trial step with the use of the \texttt{PARSEC} tables \citep[see][]{borkovits2020a}.
    \item[(iii)] The distance of the system was constrained a posteriori by minimizing the value of $\chi^2_\mathrm{SED}$.
    \item [(iv)] Finally, note that similar to our previous modeling efforts, we applied a logarithmic limb-darkening law of which the coefficients for each stars in both bands were interpolated from passband-dependent tables downloaded from the Phoebe 1.0 Legacy page\footnote{\url{http://phoebe-project.org/1.0/download}}. These tables are based on the \citet{castellikurucz04} atmospheric models and were originally implemented in former versions of the {\sc Phoebe} software \citep{2005ApJ...628..426P}.
\end{itemize}

\begin{table*}
 \centering
\caption{Orbital and astrophysical parameters of TIC\,278825952 from the joint photodynamical light curve, ETV, SED and \texttt{PARSEC} isochrone solution. Besides the usual observational system of reference related angular orbital elements ($\omega$, $i$, $\Omega$), their counterparts in the system's invariable plane related dynamical frame of reference are also given ($\omega^\mathrm{dyn}$, $i^\mathrm{dyn}$, $\Omega^\mathrm{dyn}$). Moreover, $i_\mathrm{m}$ denotes the mutual inclination of the two orbital planes, while $i_\mathrm{inv}$ and $\Omega_\mathrm{inv}$ give the position of the invariable plane with respect to the tangential plane of the sky (i.~e., in the observational frame of reference).}
 \label{tab:syntheticfit}
\begin{tabular}{@{}llll}
\hline
\multicolumn{4}{c}{orbital elements$^a$} \\
\hline
   & \multicolumn{3}{c}{subsystem}  \\
   & \multicolumn{2}{c}{A--B} & AB--C  \\
  \hline
  $P$ [days] & \multicolumn{2}{c}{$4.781023_{-0.000002}^{+0.000002}$} & $235.5499_{-0.0056}^{+0.0042}$   \\
  $a$ [R$_\odot$] & \multicolumn{2}{c}{$15.57_{-0.37}^{+0.32}$} & $230.6_{-4.9}^{+4.3}$ \\
  $e$ & \multicolumn{2}{c}{$0.00027_{-0.00008}^{+0.00020}$} & $0.00271_{-0.00107}^{+0.00233}$ \\
  $\omega$ [deg]& \multicolumn{2}{c}{$331_{-39}^{+64}$} & $144_{-35}^{+92}$ \\ 
  $i$ [deg] & \multicolumn{2}{c}{$89.93_{-0.20}^{+0.16}$} & $90.014_{-0.053}^{+0.044}$ \\
  $\mathcal{T}_\mathrm{inf}^b$ [BJD - 2400000]& \multicolumn{2}{c}{$58328.94778_{-0.00006}^{+0.00006}$} & $58401.5712_{-0.0043}^{+0.0043}$ \\
  $\Omega$ [deg] & \multicolumn{2}{c}{$0.0$} & $0.42_{-0.60}^{+0.57}$ \\
  $i_\mathrm{m}$ [deg] & \multicolumn{3}{c}{$0.48_{-0.23}^{+0.35}$} \\
  $\omega^\mathrm{dyn}$ [deg]& \multicolumn{2}{c}{$114_{-60}^{+71}$} & $142_{-97}^{+88}$ \\
  $i^\mathrm{dyn}$ [deg] & \multicolumn{2}{c}{$0.39_{-0.16}^{+0.29}$} & $0.09_{-0.04}^{+0.06}$\\
  $\Omega^\mathrm{dyn}$ [deg] & \multicolumn{2}{c}{$44_{-73}^{+48}$} & $224_{-73}^{+48}$ \\
  $i_\mathrm{inv}$ [deg] & \multicolumn{3}{c}{$90.00_{-0.05}^{+0.04}$} \\
  $\Omega_\mathrm{inv}$ [deg] & \multicolumn{3}{c}{$0.35_{-0.49}^{+0.47}$} \\
  \hline
  mass ratio $[q=m_\mathrm{sec}/m_\mathrm{pri}]$ & \multicolumn{2}{c}{$0.978_{-0.005}^{+0.002}$} & $0.339_{-0.009}^{+0.008}$ \\
  $K_\mathrm{pri}$ [km\,s$^{-1}$] & \multicolumn{2}{c}{$81.50_{-1.87}^{+1.41}$} & $12.51_{-0.11}^{+0.11}$  \\ 
  $K_\mathrm{sec}$ [km\,s$^{-1}$] & \multicolumn{2}{c}{$83.32_{-2.02}^{+1.96}$} & $37.02_{-1.03}^{+0.87}$ \\ 
  \hline  
\multicolumn{4}{c}{stellar parameters} \\
\hline
   & A & B &  C  \\
  \hline
 \multicolumn{4}{c}{Relative quantities$^c$} \\
  \hline
 fractional radius [$R/a$]  & $0.0902_{-0.0003}^{+0.0003}$ & $0.0838_{-0.0005}^{+0.0005}$  & $0.00300_{-0.00003}^{+0.00003}$ \\
 fractional flux [in \textit{TESS}-band] & $0.5119$  & $0.4333$    & $0.0503$ \\
 fractional flux [in WASP-band]& $0.5224$  & $0.4388$    & $0.0382$ \\
 \hline
 \multicolumn{4}{c}{Physical Quantities} \\
  \hline 
 $m$ [M$_\odot$] & $1.119_{-0.078}^{+0.074}$ & $1.094_{-0.075}^{+0.065}$ & $0.746_{-0.032}^{+0.034}$ \\
 $R^c$ [R$_\odot$] & $1.403_{-0.030}^{+0.037}$ & $1.306_{-0.037}^{+0.023}$ & $0.695_{-0.021}^{+0.023}$ \\
 $T_\mathrm{eff}^c$ [K]& $6261_{-69}^{+97}$ & $6229_{-71}^{+95}$ & $4894_{-88}^{+115}$ \\
 $L_\mathrm{bol}^c$ [L$_\odot$] & $2.683_{-0.109}^{+0.237}$ & $2.265_{-0.099}^{+0.205}$ & $0.250_{-0.019}^{+0.022}$ \\
 $M_\mathrm{bol}^c$ & $3.70_{-0.09}^{+0.04}$ & $3.88_{-0.09}^{+0.05}$ & $6.28_{-0.09}^{+0.08}$ \\
 $M_V^c           $ & $3.71_{-0.10}^{+0.05}$ & $3.90_{-0.10}^{+0.05}$ & $6.61_{-0.14}^{+0.13}$ \\
 $\log g^c$ [dex] & $4.190_{-0.012}^{+0.008}$ & $4.243_{-0.011}^{+0.006}$ & $4.625_{-0.008}^{+0.008}$ \\
 \hline
$\log$(age) [dex] &\multicolumn{3}{c}{$9.682_{-0.070}^{+0.078}$} \\
$[M/H]$  [dex]    &\multicolumn{3}{c}{$-0.090_{-0.207}^{+0.127}$} \\
$E(B-V)$ [mag]    &\multicolumn{3}{c}{$0.078_{-0.021}^{+0.027}$} \\
extra light $\ell_4$  [in \textit{TESS}-band]&\multicolumn{3}{c}{$0.055_{-0.006}^{+0.005}$} \\
extra light $\ell_4$  [in WASP-band]&\multicolumn{3}{c}{$0.014_{-0.008}^{+0.010}$} \\
$(M_V)_\mathrm{tot}^c$           &\multicolumn{3}{c}{$3.01_{-0.10}^{+0.05}$} \\
distance [pc]                &\multicolumn{3}{c}{$590_{-15}^{+11}$}  \\  
\hline
\end{tabular}

\textit{Notes. }{$a$: Instantaneous, osculating orbital elements, calculated for epoch $t_0=2458320.0$ (BJD); $b$: For the nearly circular orbits found here, the times of periastron passages ($\tau$) are highly uncertain, therefore, we give the inferior conjunction times (of the less massive components) for both the close and wide binaries instead; $c$: Interpolated from the \texttt{PARSEC} isochrones.}
\end{table*}

\begin{figure*}
\begin{center}
\includegraphics[width=0.49 \textwidth]{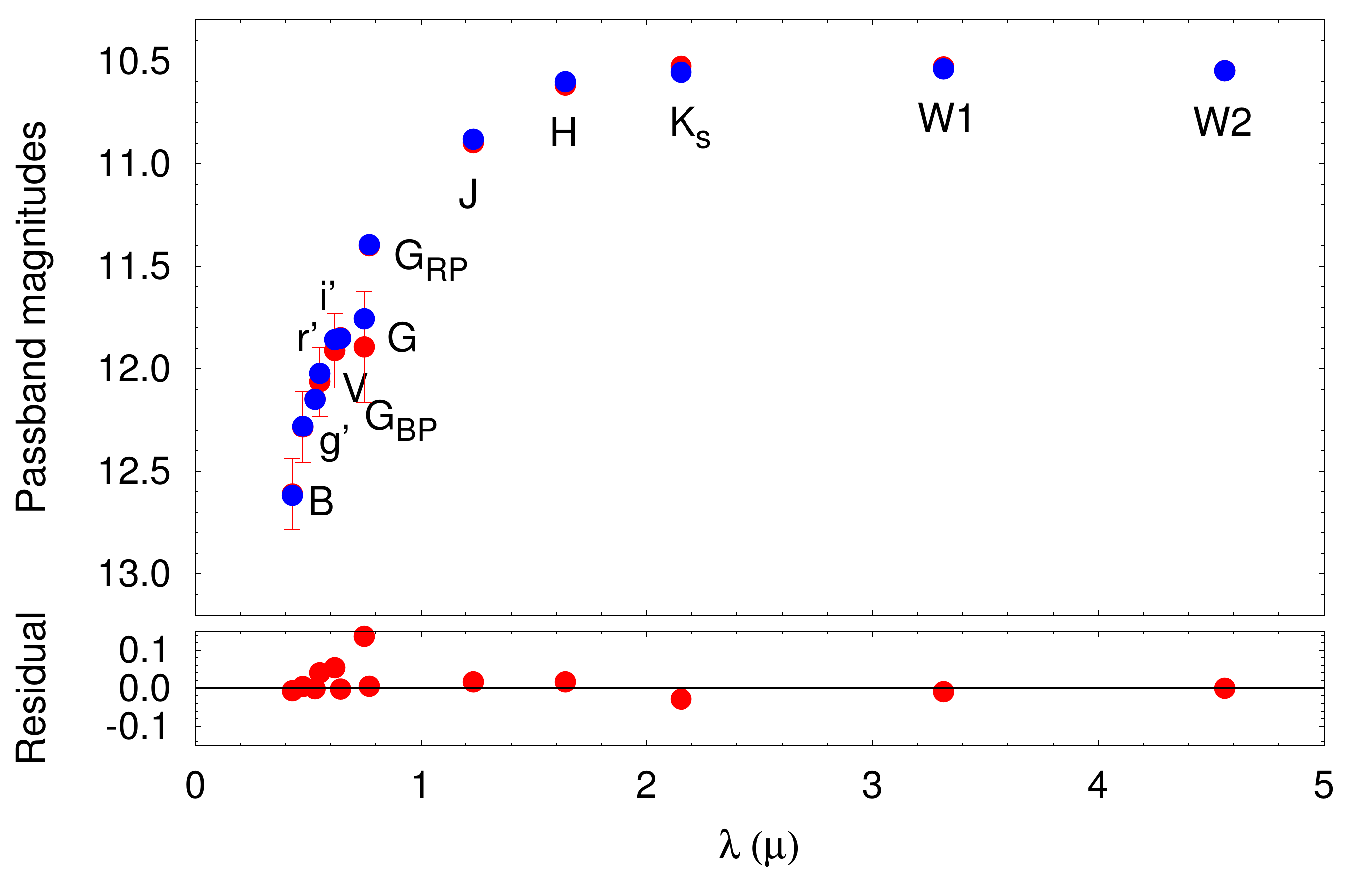}  
\includegraphics[width=0.49 \textwidth]{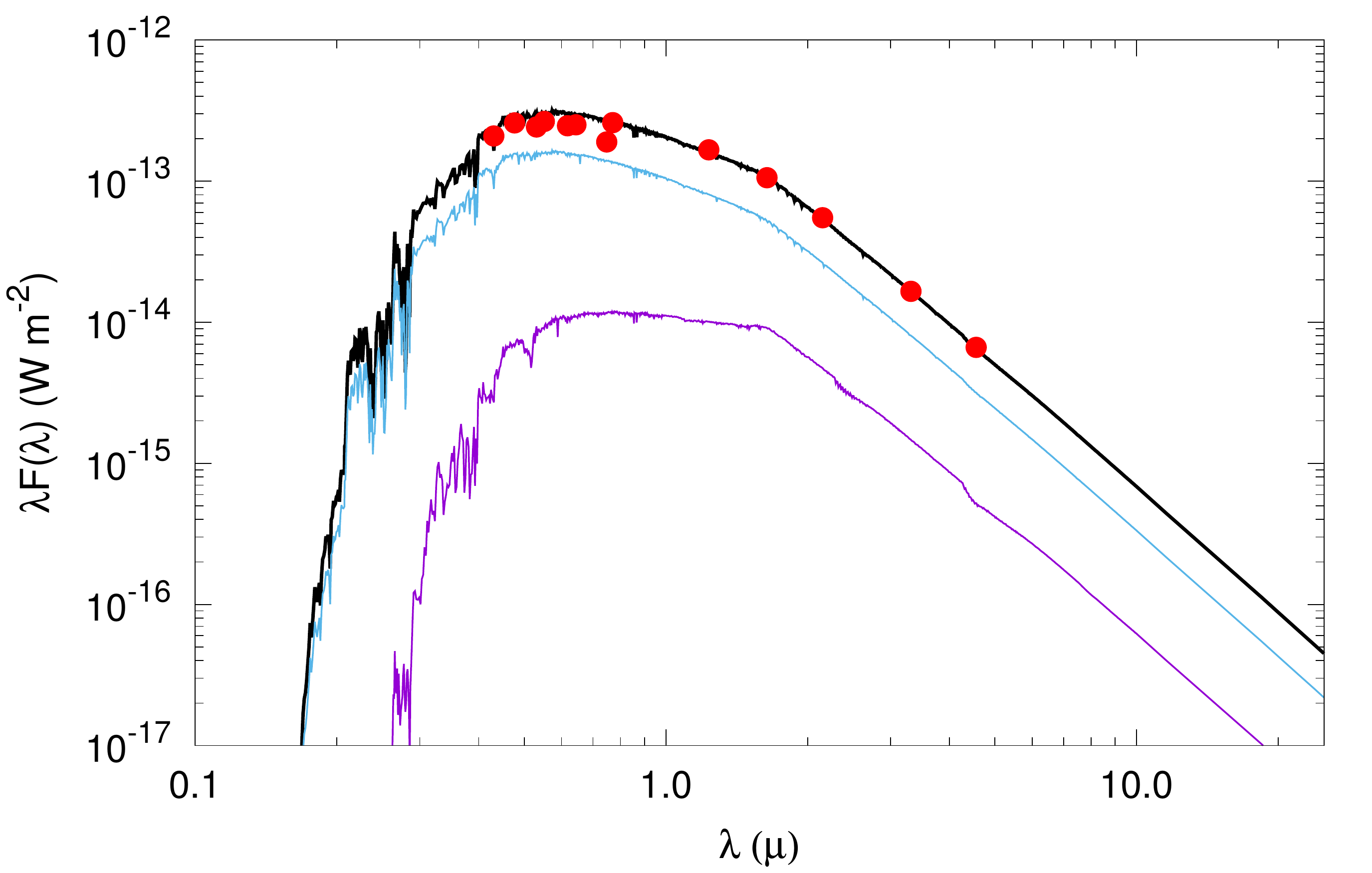}
\caption{The summed SED of the three stars of TIC\,278825952 both in the magnitude and the flux domains. The {\it left panel} displays the cataloged values of the passband magnitudes (red filled circles; tabulated in Table~\ref{tab:catalogs}) versus the model passband magnitudes derived from the absolute passband magnitudes interpolated with the use of the \texttt{PARSEC} tables (blue filled circles).  In the {\it right panel} the dereddened observed magnitudes are converted into the flux domain (red filled circles), and overplotted with the quasi-continuous summed SED for the triple star system (thick black line). This SED is computed from the \citet{castellikurucz04} ATLAS9 stellar atmospheres models (\url{http://wwwuser.oats.inaf.it/castelli/grids/gridp00k2odfnew/fp00k2tab.html}). The separate SEDs of the twin stars of the inner binary and of the less massive third component are also shown with thin green and purple lines, respectively. } 
\label{fig:sedfit} 
\end{center}
\end{figure*}  

\section{Results and discussion}
\label{sec:discussion}

The median values of the orbital and physical parameters of the system derived from the MCMC posteriors and their $1\sigma$ statistical uncertainties are summarized in Table~\ref{tab:syntheticfit}. Furthermore, the synthetic model light curves derived from the best-fit joint solution are displayed in Figs.~\ref{fig:outecl} and \ref{fig:wasp_lc}, while the corresponding ETV curves are presented in Fig.~\ref{etvs}. Finally, in the two panels of Fig.~\ref{fig:sedfit}, we illustrate the goodness of the SED-fitting part of the combined solution both in the flux and the passband magnitude domain.

According to our model, the inner binary of TIC\,278825952 consists of two almost identical, slightly evolved main sequence stars with masses of $m_\mathrm{A}=1.12_{-0.08}^{+0.07}$\,M$_\odot$, $m_\mathrm{B}=1.09_{-0.07}^{+0.08}$\,M$_\odot$ and effective temperatures of $T_\mathrm{eff,\,A}=6261_{-69}^{+97}$\,K, $T_\mathrm{eff,\,B}=6229_{-71}^{+95}$\,K. The outer tertiary component has a lower mass of $m_\mathrm{C}=0.75_{-0.03}^{+0.03}$\,M$_\odot$ and effective temperature of T$_\mathrm{eff,\,C}=4894_{-88}^{+115}$\,K. According to the SED and \texttt{PARSEC} isochrone-fitting part of the combined analysis, this $\tau=4.81_{-0.72}^{+0.94}$\,Gyr-old system most probably has a solar like metallicity of $[M/H]=-0.09_{-0.21}^{+0.13}$. The corresponding distance, obtained after taking into account the reddening of $E(B-V)=0.08_{-0.02}^{+0.03}$, is found to be $d=590_{-15}^{+11}$\,pc, which is slightly higher than, but within $2\sigma$ of, the trigonometric distance of $d_\mathrm{DR2}=561\pm8$\,pc derived from the Gaia DR2 parallax by \citet{bailer-jones2018}. 

\begin{figure*}
\begin{center}
\includegraphics[width=0.47 \textwidth]{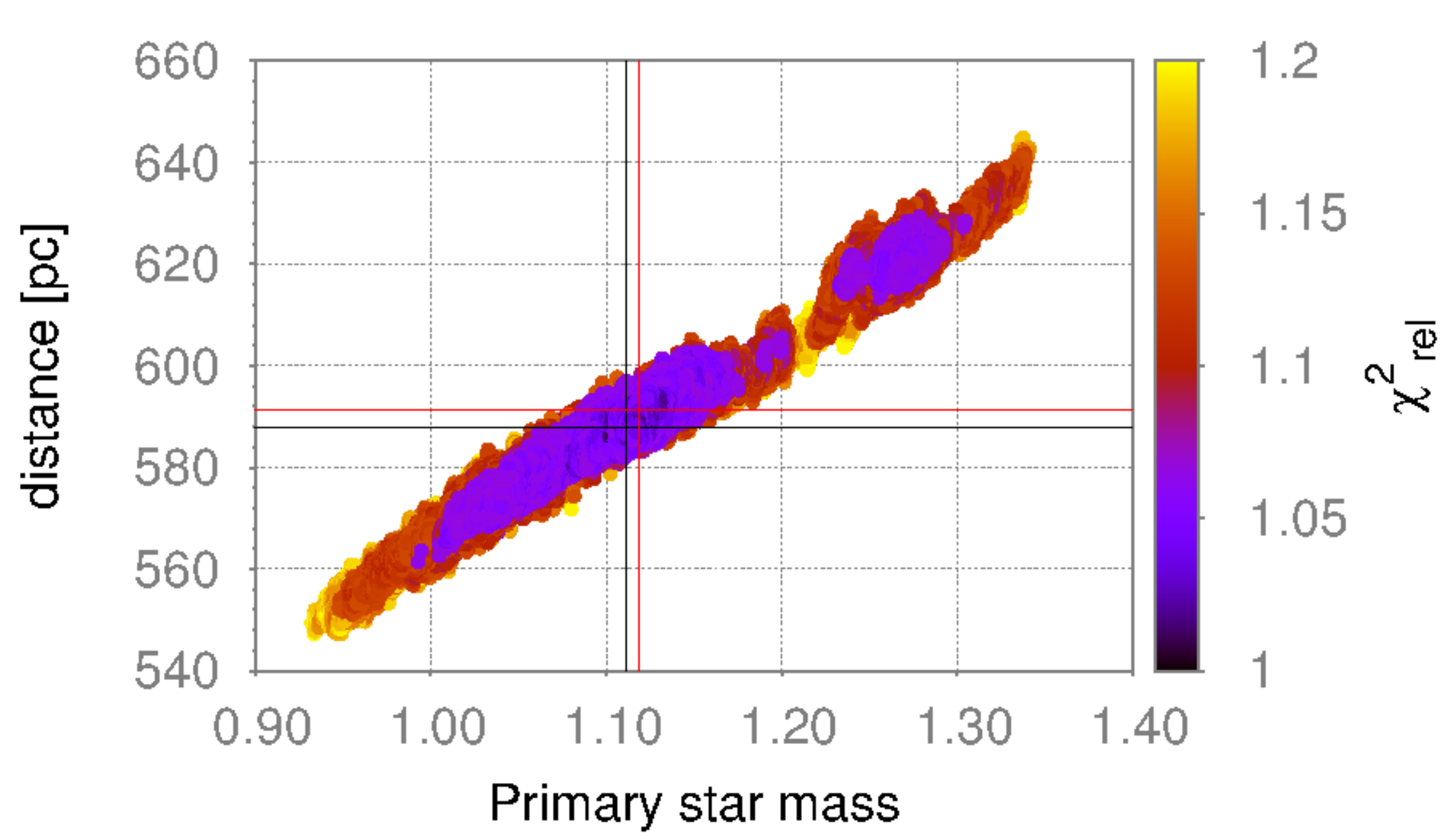}  
\includegraphics[width=0.47 \textwidth]{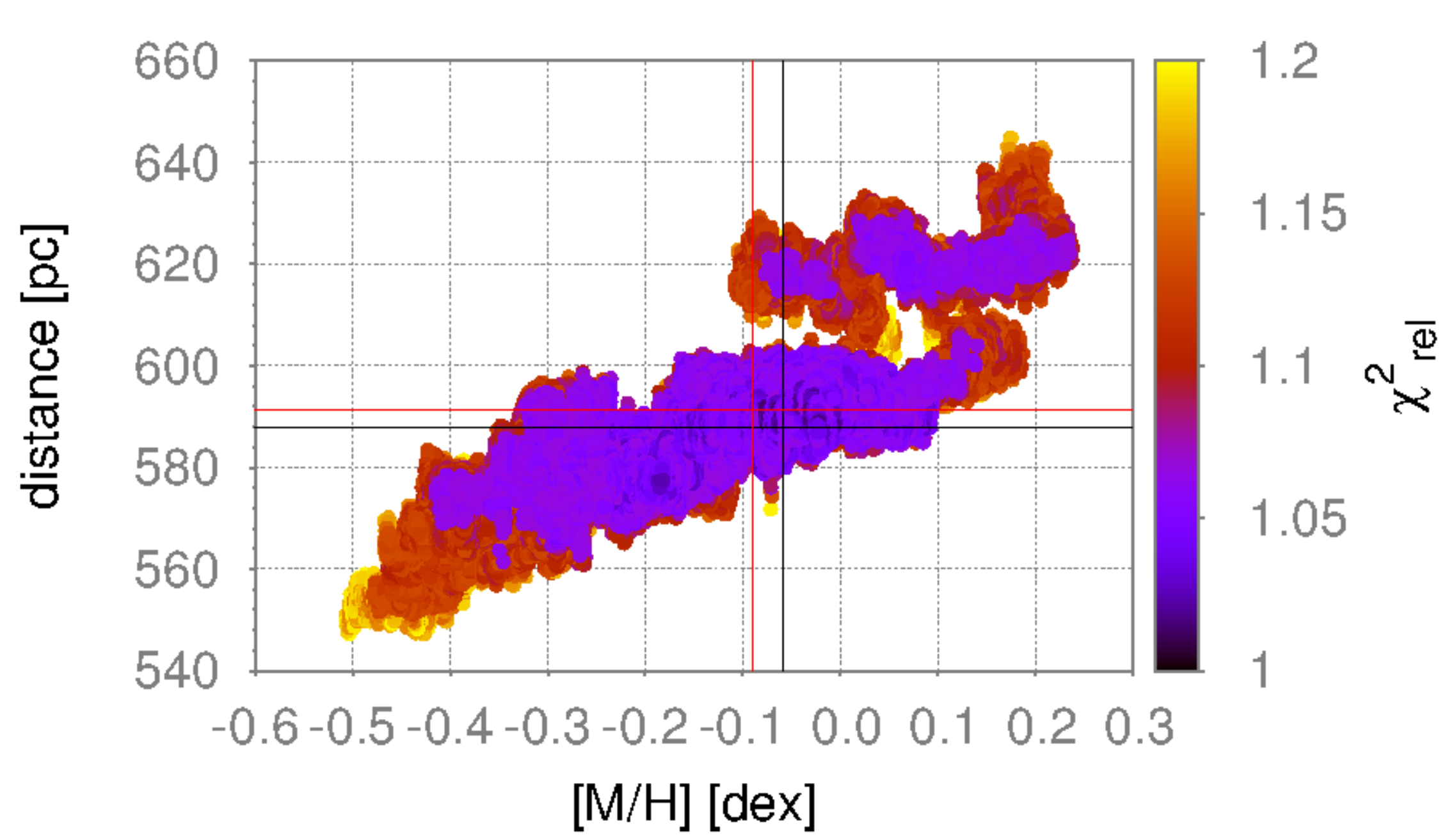}
\includegraphics[width=0.47 \textwidth]{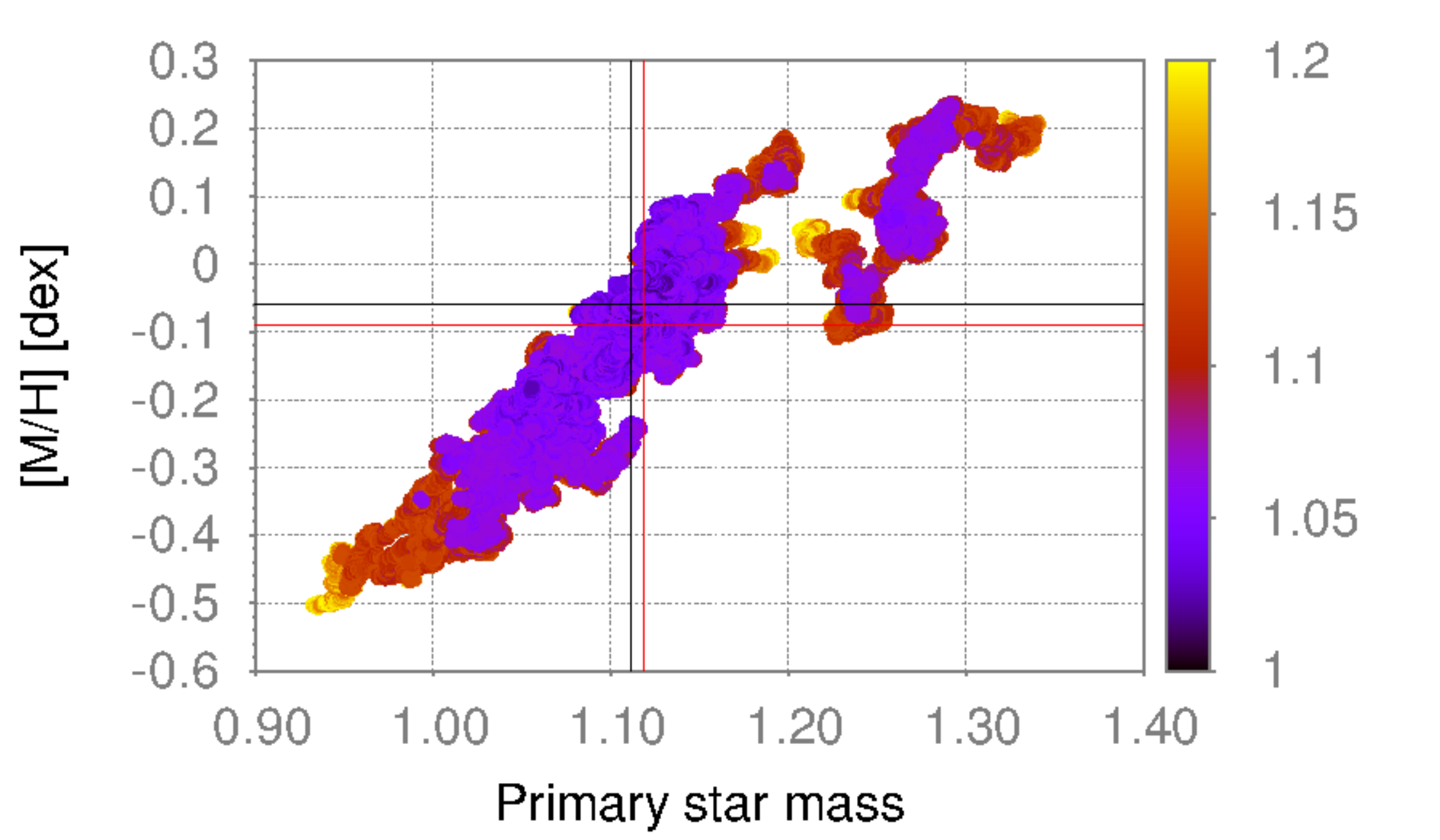}
\includegraphics[width=0.47 \textwidth]{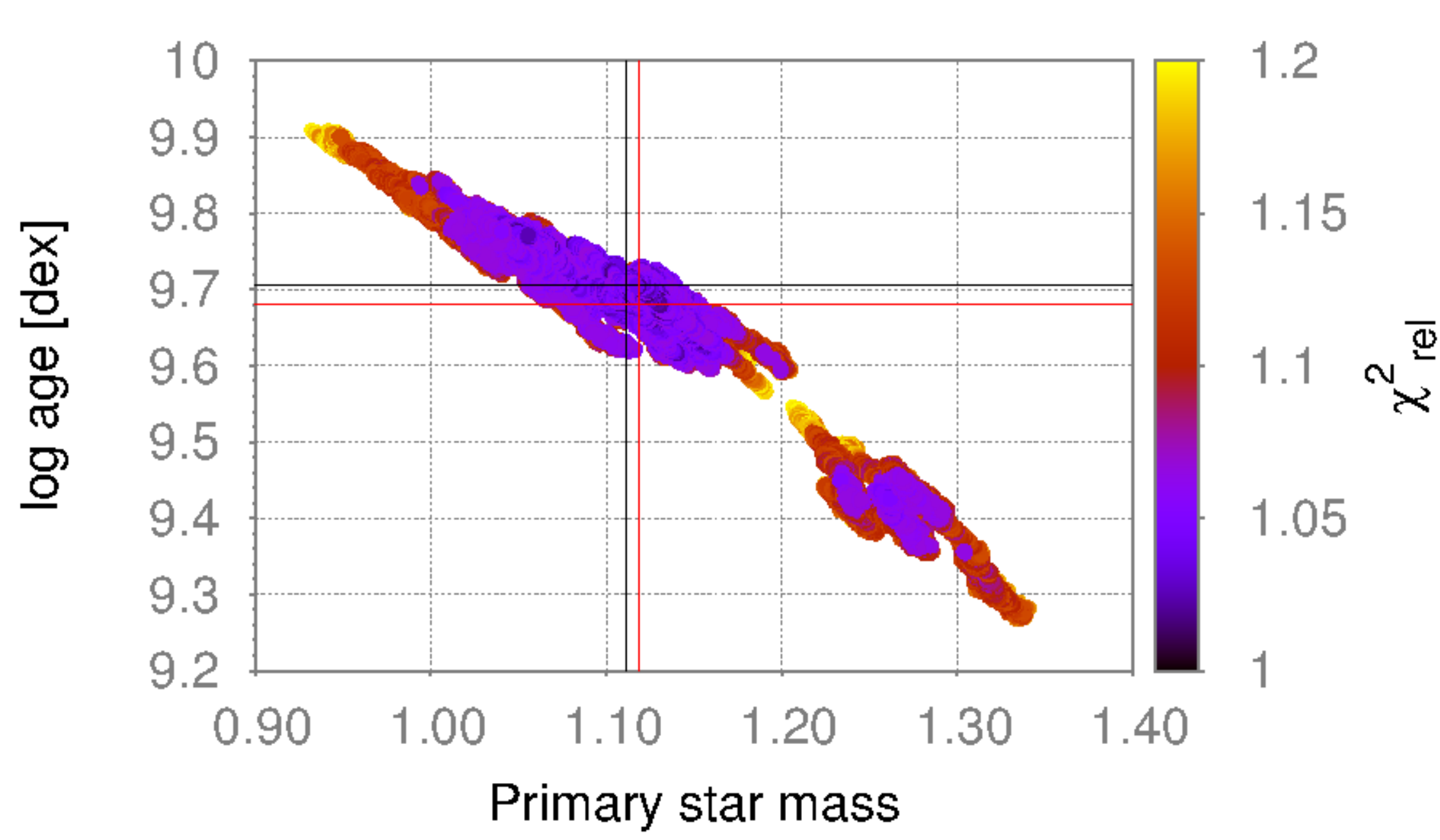}
\caption{Some correlation plots of TIC\,278825952. {\it Upper panels:} The correlations of the system distance, $d$, vs the mass of the most massive component, $m_\mathrm{A}$, and the metallicity, [$M/H$] (left and right, respectively). {\it Lower panels.} The primary mass vs. metallicity, and the logarithmic age, $\log\tau$, correlations (left and right, respectively), illustrating the natural relations between stellar mass, metallicity and age, coming from the \texttt{PARSEC} evolutionary track. The parameters belonging to the median and the best-fit values are denoted with red and black crosses. The color scale is set relative to the minimum of the $\chi^2$ value.} 
\label{fig:corrplots} 
\end{center}
\end{figure*}

Regarding this slight discrepancy, we show correlation plots of the a posteriori distributions of $m_\mathrm{A}$ and metallicity ($[M/H)$) versus photometric distance in the first row of Fig.~\ref{fig:corrplots}. Use of the Gaia DR2 distance would lead to a primary mass of about $m_\mathrm{A}\sim1\,\mathrm{M}_\odot$ and also a metallicity of $[M/H]\sim-0.3 ~{\rm to} -0.4$. Note, that both spectroscopic studies cited previously in Sect.~\ref{sec:obsdata} and listed in Table~\ref{tab:catalogs} have resulted in a metallicity in this latter range, while our photodynamical analyses evidently prefer somewhat different results. These findings show that some caution is needed in regard to those parameters of our solution which primarily depend on the evolutionary tracks (especially $m_\mathrm{A}$, $[M/H]$ and $\log\tau$). Therefore, dynamical mass determinations based on future radial velocity measurements are highly desirable for confirming our results or, helping to refine the \texttt{PARSEC} evolutionary tracks close to the terminal age main sequence stage of our stars.

Turning to the astrophysical-model independent dynamical properties, both the inner and outer orbits are very close to circular and, as the low value of mutual inclination indicates, the system is also very flat. This should make the orbital parameters of the system stable on the nuclear timescale of the stars. While the circular inner orbit for such an old and close binary is quite natural, one cannot state the same for the outer orbit. Oppositely, it is quite rare that the third, distant component of a hierarchical triple star system (even for quite compact outer orbits) would have an almost circular orbit. For example, \citet{borko16} has reported 22 triple star candidates with outer period $P_2\leq240$\,d among the eclipsing binaries in the original \textit{Kepler}-field, and only for five of them was an outer eccentricity of $e_2\leq0.1$ found. More specifically, considering only the flat, compact hierarchical triple systems with accurately known parameters tabulated in Table~5 of \citet{borkovits2020b} only one of them has outer orbit less eccentric than $e_2\sim0.2$. This only exception, HD\,181068 is similar to the present system not only in its flatness, but it also has two circular orbits \citep{borko13}. On the other hand, however, in the case of HD\,181068 the distant component is a red giant revolving on a much closer outer orbit ($P_2\sim45.5$\,d), and in such a manner, the circular outer orbit might be explained by either (i) the much stronger tidal effects during the red giant phase, or (ii) the effects of mass loss and mass transfer in the system. In the case of TIC\,278825952 we cannot find any physical reasons for circularization of the outer orbit, and therefore we suppose that it is of primordial origin. Formation scenarios for wide multiple star systems such as this one are quite uncertain, but a number of interesting idea have been put forth by e.~g. \citet{2010ApJ...710.1375K}, \citet{2016MNRAS.456.4219A}, and \citet{2017ApJ...844..103T}.

Turning to the flatness of the system, we checked the possibility of a flat, but retrograde configuration, but all the MCMC chains initiated with $\Omega_2\sim180\degr$ resulted in significantly higher $\chi^2$ values, i.e., we found that $(\chi^2_\mathrm{min})_\mathrm{retrograde}\sim1.25\times(\chi^2_\mathrm{min})_\mathrm{prograde}$. Therefore, we conclude that the system most probably has a flat, prograde configuration.

In Table~1 of \citet{borkovits2020b} we listed the inner and outer periods of all the 17 triply eclipsing triple stars having precisely known inner and outer orbital periods. TIC\,278825952 now joins this small group of triple stars. Its outer period ($P_2$) places it in the tenth position (in increasing order). From a dynamical point of view, it is one of the most relaxed systems, especially considering: (i) the relatively high period ratio of $P_2/P_1$=49.3 and low outer mass ratio of $q_2=0.34$, the two parameters which basically set the amplitude of the dynamical perturbations, and also (ii) the doubly circular, coplanar configuration of the inner and outer orbits.  These two effects render the lowest order (quadruple) perturbative terms to be nearly zero \citep[see, e.~g.][for the `apse-node' and the $P_2$-timescale perturbations, respectively]{soderhjelm82,borko03}. Furthermore, the nearly equal masses of the inner binary stars ($q_1=0.98$) also substantially reduce the amplitudes of the next, octuple-order perturbations, which disappear when the inner mass ratio tends toward unity  \citep[see again][for the two above classes of third-body perturbations]{soderhjelm82,borko15}.

\section{Summary}
\label{Sect:Summary}

In this paper, we have reported the discovery and  first comprehensive analysis of the triply eclipsing hierarchical triple star TIC\,278825952 observed by the \textit{TESS} spacecraft almost continuously for $\sim$11 months in its SCVZ. The space-borne observations cover more than one outer orbital cycle, including third-body eclipsing events around two consecutive inferior and one superior conjunctions of the distant, third star, allowing the accurate determination of the dynamical and astrophysical parameters of the system. In order to obtain these parameters with the highest available accuracy, we carried out a joint photodynamical analysis that included not only the above mentioned \textit{TESS} light curve, but also earlier archival ground-based WASP photometry, the ETVs extracted from these observations and, furthermore, the SED and theoretical \texttt{PARSEC} isochrones. Note, however, that due to the lack of radial velocity observations, we were unable to carry out a fully model-independent, purely dynamical study of the stellar masses.

Our comprehensive analysis revealed that TIC\,278825952 consists of a near-twin pair of slightly evolved main sequence stars on a circular orbit with a lower mass outer companion that is also on a circular orbit. The system is one of the few members of the currently known class of hierarchical triple star systems exhibiting outer third-body eclipses in a highly coplanar configuration. Nevertheless, it is unique in that it has the most inherently circular outer orbit among them, raising a question about the origin of the low eccentricity of its wide orbit. For a system with such age and constituent stars, this is unexpected and we could not find any physical reasons that explain the highly circular nature of the outer orbit, so we propose that it most probably represents its primordial configuration.

Because of its fortuitous location in the SCVZ, TIC\,278825952 is also scheduled to be observed in all the 13 sectors of the first year of the \textit{TESS} extended mission. According to our model the forthcoming outer eclipsing events are expected to be observed during Sectors 29, 33, and 38 (and hopefully will be), leading to a more refined photodynamical model.

\section*{Acknowledgements}

T.\,M. and T.\,B. acknowledge the financial support of the Hungarian National Research, Development and Innovation Office -- NKFIH Grant KH-130372.

This paper includes data collected by the \textit{TESS} mission. Funding for the \textit{TESS} mission is provided by the NASA Science Mission directorate. Some of the data presented in this paper were obtained from the Mikulski Archive for Space Telescopes (MAST). STScI is operated by the Association of Universities for Research in Astronomy, Inc., under NASA contract NAS5-26555. Support for MAST for non-HST data is provided by the NASA Office of Space Science via grant NNX09AF08G and by other grants and contracts.

This work has made use  of data  from the European  Space Agency (ESA)  mission {\it Gaia}\footnote{\url{https://www.cosmos.esa.int/gaia}},  processed  by  the {\it   Gaia}   Data   Processing   and  Analysis   Consortium   (DPAC)\footnote{\url{https://www.cosmos.esa.int/web/gaia/dpac/consortium}}.  Funding for the DPAC  has been provided  by national  institutions, in  particular the institutions participating in the {\it Gaia} Multilateral Agreement.

This publication makes use of data products from the Wide-field Infrared Survey Explorer, which is a joint project of the University of California, Los Angeles, and the Jet Propulsion Laboratory/California Institute of Technology, funded by the National Aeronautics and Space Administration. 

This publication makes use of data products from the Two Micron All Sky Survey, which is a joint project of the University of Massachusetts and the Infrared Processing and Analysis Center/California Institute of Technology, funded by the National Aeronautics and Space Administration and the National Science Foundation.

We  used the  Simbad  service  operated by  the  Centre des  Donn\'ees Stellaires (Strasbourg,  France) and the ESO  Science Archive Facility services (data  obtained under request number 396301).

\section*{Data availability}

The \textit{TESS} data underlying this article were accessed from MAST (Barbara A. Mikulski Archive for Space Telescopes) Portal (\url{https://mast.stsci.edu/portal/Mashup/Clients/Mast/Portal.html}), including the data products found in the bulk download website (http://archive.stsci.edu/tess/bulk\_downloads/bulk\_downloads\_ffi-tp-lc-dv.html). Part of the data were derived from sources in public domain as given in the respective footnotes. The derived data generated in this research and the code used for the photodynamical analysis will be shared on reasonable request to the corresponding author.


\bibliographystyle{mnras}
\bibliography{example} 


\bsp	
\label{lastpage}
\end{document}